\documentclass[aps,prd,longbibliography,twocolumn,superscriptaddress]{revtex4-2}
\usepackage{graphicx}
\usepackage{subfigure}
\usepackage{float}
\usepackage{amsmath}
\usepackage{amsfonts}
\usepackage{color}
\usepackage{bm}
\usepackage{bbm}
\usepackage{txfonts}
\usepackage{array}
\usepackage{amssymb}
\usepackage[english]{babel}
\usepackage{soul}
\usepackage{placeins}
\usepackage{orcidlink}   

\hypersetup{
    colorlinks=true,
    linkcolor=red,
    urlcolor=magenta,
    citecolor=blue,
    anchorcolor=blue
}

\begin{document}

\title{Spin asymmetry in vacuum pair production under polarized electric fields}
\author{Dan-Yang Xu}
\author{Jian Mao}

\author{Mamat Ali Bake\,\orcidlink{0000-0001-6464-4591}}
\email[Correspondence Author. email:]{mabake@xju.edu.cn}
\affiliation{Xinjiang Key Laboratory of Solid State Physics and Devices, School of Physical Science and Technology, \href{https://ror.org/059gw8r13}{Xinjiang University}, Urumqi 830017, People's Republic of China}

\author{Orkash Amat\,\orcidlink{0000-0003-3230-7587}}
\email[Correspondence Author. email:] {orkashamat@nju.edu.cn}
\affiliation{School of Astronomy and Space Science, \href{https://ror.org/01rxvg760}{Nanjing University}, Nanjing 210023, People's Republic of China}
\affiliation{State Key Laboratory of Radio Astronomy and Technology, \href{https://ror.org/058pyyv44}{National Astronomical Observatories}, \href{https://ror.org/034t30j35}{Chinese Academy of Sciences}, Beijing 100101, People's Republic of China}

\begin{abstract}
We study spin-resolved electron--positron pair production from the quantum vacuum in spatially homogeneous, time-dependent electric fields with linear, elliptical, and circular polarizations. The equal-time Dirac--Heisenberg--Wigner formalism is employed to compute the final momentum distributions in two distinct frequency regimes, for both
unchirped and chirped field configurations. In the high-frequency regime, characterized by a reference Keldysh parameter, linearly polarized fields yield identical particle yields for the two spin projections considered, whereas elliptical and circular polarizations generate pronounced spin-resolved population imbalances near the dominant multiphoton
resonance. Frequency chirping broadens the resonant spectral structures and extends the momentum range over which a sizable normalized spin difference persists. In the low-frequency regime, the pair-production spectra are governed by quantum tunneling and temporal interference effects. Although the peak spin asymmetry in this regime is substantially weaker than in the high-frequency case, the normalized differential spectra still exhibit sign-changing interference fringes. We find that the spin-asymmetry degree exceeds 90$\%$ for circular polarization in the high-frequency regime, while it saturates at approximately 20$\%$ in the low-frequency regime. Given the purely electric background field, we interpret these results in terms of spin-projected Wigner components and the handed time dependence of rotating electric fields, while clearly distinguishing normalized spectral differences from local or integrated polarization observables.	
\end{abstract}

\maketitle

\section{Introduction}
The spontaneous creation of electron-positron ($e^+e^-$) pairs from the quantum vacuum in a strong external electric field (Schwinger effect) is a fundamental prediction of quantum electrodynamics (QED)~\cite{Schwinger1951}. In the tunneling regime, the external field destabilizes the vacuum across the electron-positron mass gap $2m$, where $m$ is the electron mass~\cite{Sauter1931}. For a constant, subcritical electric field $E$, the leading spinor-QED pair-production rate per unit volume in natural units is $\Gamma\simeq (|e|E)^2/(4\pi^3)\exp[-\pi m^2/(|e|E)]$. Restoring constants gives $\Gamma\simeq |e|^2E^2/(4\pi^3\hbar^2c)\exp(-\pi E_{\mathrm{cr}}/E)$, where $E_{\mathrm{cr}}=m^2c^3/(|e|\hbar)\simeq1.32\times10^{16}\,\mathrm{V/cm}$. This leading term is exponentially suppressed below the critical field; rapidly varying fields can also produce pairs through perturbative multiphoton channels. The corresponding plane-wave intensity scale is $I_{\mathrm{cr}}\sim 2.3\times10^{29}\,\mathrm{W/cm^2}$ when $I=c\epsilon_0E^2/2$ is expressed using the peak field~\cite{dunne2005,fedotov2023}. These field scales remain far beyond those directly attained by present laser facilities, making the observation of Schwinger vacuum pair production an outstanding challenge~\cite{fedotov2023,yoon2021}.

Although the Schwinger field remains out of experimental reach, advances in ultraintense laser technology have renewed interest in strong-field QED~\cite{dipiazza2012}. Facilities such as the Extreme Light Infrastructure (ELI)~\cite{turcu2016,gales2018}, the Station of Extreme Light (SEL)~\cite{shen2018}, and the Center for Relativistic Laser Science (CoReLS)~\cite{sung2017} motivate studies of peak intensities around $10^{23}\,\mathrm{W/cm^2}$ and proposed higher-intensity configurations~\cite{yoon2021,sung2017,gales2018}. Dynamically assisted Schwinger pair production, in which a strong, slowly varying field is supplemented by a weaker, rapidly varying component, provides a theoretical mechanism for enhancing subcritical pair yields~\cite{schutzhold2008}. Whether a particular configuration is experimentally accessible requires a separate assessment of its frequency, focal geometry, and field strength.

Advances in high-intensity laser facilities and numerical methods have stimulated research on Schwinger pair production. Two related objectives are to enhance pair yields and to identify signatures of the underlying quantum dynamics.
The first objective is to enhance pair yields in subcritical electric fields~\cite{Schwinger1951,heisenberg1936}. To this end, strategies such as the dynamically assisted Schwinger mechanism~\cite{schutzhold2008}, chirped electric fields~\cite{dumlu2010}, and optimized field pulses have been
introduced~\cite{kohlfurst2013}. In this direction, studying pair production through momentum distribution functions has proven to be a particularly effective approach, revealing a variety of important signatures in the enhanced yields, including multiphoton resonances and temporal-interference fringes. The second objective is to understand the quantum dynamics and identify informative observables. This line of inquiry not only deepens the theoretical understanding of the Schwinger mechanism but also provides a vital basis for interpreting future experimental data.

Beyond particle yields and momentum spectra, the spin degrees of freedom of created pairs encode information about their interaction with external fields. Electrons and positrons are spin-$1/2$ fermions governed by the Dirac equation. Time-dependent rotating fields can couple the spin and momentum sectors and produce unequal populations in a specified spin basis. Linear, circular, and elliptical polarizations modify the momentum distributions~\cite{xie2012,li2019}; spin asymmetry additionally depends on the chosen spin projector and the field history. The Dirac-Heisenberg-Wigner (DHW) formalism, quantum kinetic theory (QKT), and generalized quantum two-level models (TLMs) provide complementary descriptions, with distinct approximation requirements and conventions for spin-resolved quantities~\cite{amat2025spin}.

In recent years, a large body of research has demonstrated that spin effects play a crucial role in vacuum pair production. Strobel \emph{et al.} showed that Schwinger pairs generated by a time-varying rotating electric field exhibit a nonuniform spin distribution~\cite{strobel2015}. Blinne pointed out that, for a constant rotating field, the populations of the two spin states are unequal~\cite{blinne2016}. Hu \emph{et al.} studied electron-positron pairs produced by circularly polarized fields with two different frequencies and a variable time delay, finding helical structures in both the spin-up and spin-down momentum distributions, and that the degree of spin asymmetry can be significantly altered by adjusting the delay time~\cite{hu2024}. Jiang \emph{et al.} investigated multiperiod elliptically polarized fields and found that the spin asymmetry increases with the Keldysh adiabaticity parameter; furthermore, by analyzing the spin states of pairs produced in circularly polarized electric fields, they showed that the helical momentum distribution is a special case of the spin-resolved momentum distribution, obtained by aligning the spin quantization axis with the particle momentum~\cite{li2026}. In one of our previous studies based on the DHW formalism, we investigated two-color counter-rotating electric fields and found that vacuum pair production in a pure electric field exhibits spin dependence and is highly sensitive to changes in the field parameters; for asymmetric field configurations, increasing the field asymmetry and the chirp parameter leads to a reversal of the spin asymmetry~\cite{chen2025}. At the same time, other studies have indicated that combining different frequency chirps with electric field polarization can modify the pair yields and momentum distributions~\cite{jangir2026}. These studies already establish polarization-dependent spin asymmetry. Here we examine its dependence on ellipticity and quadratic chirp in a Gaussian pulse, comparing two specified frequency regimes and distinguishing spectral asymmetry indicators from polarization.

In this work, we investigate spin-resolved Schwinger pair production in spatially homogeneous, time-dependent electric fields with variable ellipticity and quadratic chirp. Using the equal-time DHW formalism, we calculate momentum spectra in high- and low-frequency regimes. The displayed spin-up and spin-down channels coincide for linear polarization, whereas rotating fields produce unequal populations in the specified final-state basis. In the high-frequency chirped circular case, the displayed spin-down channel dominates and the peak-asymmetry indicator can exceed $0.9$ in magnitude. In the low-frequency parameter scan, this particular indicator remains below approximately $0.18$, while the momentum-dependent spin difference retains sign-changing interference structures. These statements concern distributions on the $q_z=0$ plane and their maxima; momentum-integrated polarization requires integration over the full three-dimensional spectrum. The calculation therefore characterizes spin-selective spectral dynamics without establishing a macroscopic beam polarization or an experimental production rate.

The paper is organized as follows. Section~\ref{field} specifies the background field, and Sec.~\ref{DHWformalism} presents the equal-time DHW formalism and spin observables. Sections~\ref{high} and \ref{low} discuss the high- and low-frequency momentum spectra and spin asymmetry indicators. We summarize the results in Sec.~\ref{sec:conclusion}. Appendix~\ref{APPENDIX} gives the corresponding quantum kinetic description and its relation to the DHW variables.
 
\section{Model of background field}\label{field}
We investigate the effects of a polarized electric field on the spin asymmetry of created pairs under various chirp conditions in both the high- and low-frequency regimes, using the DHW formalism. We consider a spatially homogeneous, time-dependent chirped electric field with arbitrary polarization,
\begin{equation}\label{1}
\mathbf{E}(t) = \frac{E_0}{\sqrt{1+\delta^2}} \exp\!\left(-\frac{t^2}{2\tau^2}\right)
\begin{bmatrix}
    \cos(\omega t + bt^2) \\
    \delta \sin(\omega t + bt^2) \\
    0
\end{bmatrix},
\end{equation}
where $\mathbf E(t)$ is the electric-field vector, $E_0/\sqrt{1+\delta^2}$ is its maximum magnitude, $\tau$ is the Gaussian envelope width, and $\omega$ is the carrier frequency at $t=0$. The ellipticity satisfies $|\delta|\leq1$: $\delta=0$, $0.5$, and $1$ give linear, elliptical, and circular polarization, respectively. We choose positive $\delta$, with the field rotating from $+x$ toward $+y$ when its phase increases; reversing $\delta$ reverses this sense of rotation. The phase is $\phi(t)=\omega t+bt^2$, so its instantaneous angular frequency is $\omega_{\mathrm{eff}}(t)=\dot\phi(t)=\omega+2bt$. Thus $b$ has dimensions of $m^2$, and chirp values are quoted through $\beta=b\tau/\omega$. For every $|\delta|\leq1$, $|\mathbf E(t)|\leq E_0\exp[-t^2/(2\tau^2)]/\sqrt{1+\delta^2}$, with equality at $t=0$. A quadratic chirp changes the field history but does not displace the global maximum of the field magnitude.

We compare two parameter sets. In the high-frequency case, $E_0=0.1E_{\mathrm{cr}}$, $\omega=0.9\,m$, and $\tau=100\,m^{-1}$; in the low-frequency case, $E_0=0.5\sqrt{2}E_{\mathrm{cr}}$, $\omega=0.05\,m$, and $\tau=100\,m^{-1}$. Because of the normalization in Eq.~\eqref{1}, the actual peak field changes with $\delta$. The factor $1/\sqrt{1+\delta^2}$ approximately fixes the cycle-averaged squared field for a slowly varying envelope; it does not fix the peak field.

A field of the form in Eq.~\eqref{1} can approximate the electric background near an electric antinode of a suitable standing wave formed by counterpropagating laser pulses. Magnetic-field cancellation occurs locally and does not make the whole standing wave purely electric or spatially homogeneous. Here we prescribe a homogeneous electric field and set $\mathbf B=0$; spatial gradients, focusing, depletion, and backreaction are omitted. Applying this approximation to a laser configuration requires checking that spatial variations across the pair-formation region are negligible.

Throughout this paper, $\hbar=c=1$, and quantities are expressed in units of the electron mass $m$. Frequencies and momenta have units of $m$, time has units of $m^{-1}$, and $b$ has units of $m^2$. In the equations of motion, $e$ denotes the signed electron charge; $E_{\mathrm{cr}}=m^2/|e|$ and regime parameters use $|e|$. We use the metric $g_{\mu\nu}=\mathrm{diag}(1,-1,-1,-1)$ and temporal gauge $A_0=0$, with $\mathbf E=-\dot{\mathbf A}$. We choose $\mathbf A(+\infty)=0$, so the asymptotic electron kinetic momentum equals $\mathbf q$; $\mathbf A(-\infty)$ need not vanish for a pulse with nonzero time integral.

\section{ Dirac-Heisenberg-Wigner formalism}\label{DHWformalism}
The DHW formalism describes Schwinger pair production in prescribed electromagnetic backgrounds~\cite{strobel2015,hebenstreit2011particle,ababekri2019effects,li2021study,osman2023efficient,bake2024enhanced,Bake2025,Bake2026,kohlfurst2018phase,kohlfurst2014effective}. It provides phase-space information, including particle distributions and spin-dependent Wigner components. We work in the external-field approximation with vacuum initial conditions. The spinor-QED Lagrangian is~\cite{hebenstreit2011particle,ababekri2019effects,li2021study,osman2023efficient,bake2024enhanced,Bake2025,Bake2026,kohlfurst2018phase,kohlfurst2014effective,kohlfurst2016effect,aleksandrov2016electron,aleksandrov2017momentum,aleksandrov2020pair,hebenstreit2009momentum,dumlu2010schwinger,dumlu2010stokes,dumlu2011interference,olugh2019pair,kohlfurst2015electron}
\begin{widetext}
\begin{equation}\label{2}
\begin{aligned}
\mathcal{L}(\Psi, \bar{\Psi}, A) = \frac{1}{2} \left( i\bar{\Psi}\gamma^\mu \mathcal{D}_\mu \Psi - i\bar{\Psi}\mathcal{D}_\mu^\dagger \gamma^\mu \Psi \right) - m\bar{\Psi}\Psi - \frac{1}{4} F_{\mu\nu} F^{\mu\nu},
\end{aligned}
\end{equation}
\end{widetext}
where $\mathcal D_\mu=\partial_\mu+ieA_\mu$ and $\mathcal D_\mu^\dagger=\overleftarrow\partial_\mu-ieA_\mu$ denote right- and left-acting covariant derivatives. The prescribed potential is $A_\mu=(A_0,-\mathbf A)$, $\gamma^\mu$ are the Dirac matrices, and $\bar\Psi=\Psi^\dagger\gamma^0$ is the Dirac adjoint of the fermion field. Electrons and positrons are excitations of the same quantized field. The Euler--Lagrange equations give the Dirac equation and its adjoint:
\begin{equation}\label{3}
\begin{aligned}
\left( \mathrm{i} \gamma^\mu \partial_\mu - e \gamma^\mu A_\mu - m \right) \Psi = 0,
\end{aligned}
\end{equation}
and the adjoint equation:
\begin{equation}\label{4}
	\begin{aligned}
		\bar{\Psi} \left( \mathrm{i} \overleftarrow{\partial_\mu} \gamma^\mu + e \gamma^\mu A_\mu + m \right) = 0.
	\end{aligned}
\end{equation}
 Next, we use the commutator of two Dirac fields to construct the covariant density operator in the Heisenberg picture:
 \begin{equation}\label{5}
 	\begin{aligned}
 		{\hat {\cal C}_{\alpha \beta }}(r,s) = U(A,r,s)\left[ {{{{\rm{\bar \Psi }}}_\beta }(r - s/2),{{\rm{\Psi }}_\alpha }(r + s/2)} \right],
 	\end{aligned}
 \end{equation}
where $r$ and $s$ are the center and relative four-coordinates. The field commutator by itself transforms under the local electromagnetic $U(1)$ gauge symmetry; the Wilson line makes the combined density operator gauge invariant:
\begin{equation}\label{6}
\begin{aligned}
U(A,r,s)=\exp\!\left[ie\int_{-1/2}^{1/2}d\xi\,s^\mu A_\mu(r+\xi s)\right].
\end{aligned}
\end{equation}
For an Abelian classical background, no non-Abelian path ordering is required. In the Hartree approximation, the electromagnetic field is treated as a prescribed c-number function~\cite{Bake2025,vasak1987quantum}. Quantum fluctuations of that field are neglected, but its time dependence is retained. The covariant Wigner operator is the Fourier transform of the density operator:
\begin{equation}\label{7}
\begin{aligned}
\hat{\mathcal W}_{\alpha\beta}(r,p)
=\frac12\int d^4s\,e^{ip_\mu s^\mu}\hat{\mathcal C}_{\alpha\beta}(r,s).
\end{aligned}
\end{equation}
The vacuum expectation value of the covariant Wigner operator is the Wigner function:
\begin{equation}\label{8}
	\begin{aligned}
		\mathbbm{W}(r, p)  =\langle\Phi|\hat{\mathcal{W}}(r, p) |\Phi\rangle.
	\end{aligned}
\end{equation}
The covariant Wigner function is a $4\times4$ matrix in Dirac space. Expanding it in the complete basis $\{\mathbbm1,\gamma_5,\gamma^\mu,\gamma^\mu\gamma_5,\sigma^{\mu\nu}\}$, with $\sigma^{\mu\nu}=i[\gamma^\mu,\gamma^\nu]/2$, gives sixteen real coefficient functions~\cite{,vasak1987quantum,bialynicki1991phasespace}. We retain the tensor normalization used in Ref.~\cite{amat2025spin}, so the coefficient multiplying $\sigma^{\mu\nu}$ has no additional factor of $1/2$:
\begin{equation}\label{9}
	\begin{aligned}
		\mathbbm{W} = \frac{1}{4} \left( \mathbbm{1} \mathbb{S} + \textrm{i} \gamma_5
		\mathbbm{P} + \gamma^{\mu} \mathbbm{V}_{\mu} + \gamma^{\mu} \gamma_5
		\mathbbm{A}_{\mu} + \sigma^{\mu \nu} \mathbbm{T}_{\mu \nu} \right).
	\end{aligned}
\end{equation}
The coefficients $\mathbb S$, $\mathbb P$, $\mathbb V_\mu$, $\mathbb A_\mu$, and $\mathbb T_{\mu\nu}$ are the scalar, pseudoscalar, vector, axial-vector, and antisymmetric tensor components, respectively~\cite{bialynicki1991phasespace}. They encode mass, pseudoscalar, current, axial, and dipole densities. The spatial tensor components describe magnetic dipole density, while time-space components describe electric dipole density. An axial Wigner component alone is not an arbitrary single-particle spin probability or an orbital angular momentum eigenvalue~\cite{zdf5-w21r}. After energy integration, the equal-time Wigner matrix obeys~\cite{hebenstreit2011particle,bialynicki1991phasespace,amat2025spin}
\begin{widetext}
\begin{equation}\label{10}
\begin{aligned}
D_t\mathbb W={}&-\frac12\mathbf D_{\mathbf x}\!\cdot
[\gamma^0\bm\gamma,\mathbb W]
-im[\gamma^0,\mathbb W]-i\mathbf P\!\cdot\{\gamma^0\bm\gamma,\mathbb W\},
\end{aligned}
\end{equation}
\end{widetext}
where $D_t$, $\mathbf D_{\mathbf x}$, and $\mathbf P$ are pseudodifferential operators. For a spatially homogeneous electric field with $\mathbf B=0$, they reduce to $D_t=\partial_t+e\mathbf E(t)\cdot\nabla_{\mathbf p}$, $\mathbf D_{\mathbf x}=\nabla_{\mathbf x}$, and $\mathbf P=\mathbf p$. The equal-time matrix $\mathbb W$ and its coefficient functions are obtained by integrating the corresponding covariant quantities over energy:
\begin{equation}\label{11}
\begin{aligned}
{\mathbbm w}(\mathbf x,\mathbf p,t)=\int\frac{dp_0}{2\pi}\,\mathbbm W(r,p).
\end{aligned}
\end{equation}
For the spatially homogeneous background in Eq.~\eqref{1}, the method of characteristics converts the component equations into ordinary differential equations at fixed canonical momentum. Vacuum initial data leave at most ten nonzero equal-time components~\cite{kohlfurst2015electron,hebenstreit2011schwinger}:
\begin{equation}\label{12}
\begin{aligned}
\mathbbm w_{10}(\mathbf q,t)
=(\mathbbm s,\mathbbm{v},
\mathbbm{a},\mathbbm{t})^T(\mathbf q,t),
\end{aligned}
\end{equation}
\begin{equation}\label{13}
	\begin{aligned}
		\dot{\mathbbm{w}}_{10}(\mathbf{q}, t) = \mathcal{M}(\mathbf{q}, t) \mathbbm{w}_{10}(\mathbf{q}, t),
	\end{aligned}
\end{equation}
where $\mathbf q$ is the canonical momentum, $\mathbf p(t)=\mathbf q-e\mathbf A(t)$ is the kinetic momentum, and $\mathcal M$ is a $10\times10$ matrix determined by $\mathbf p(t)$ and $m$. We define the tensor vector by $\mathbbm t_i=\mathbbm T_{0i}-\mathbbm T_{i0}=2\mathbbm T^{i0}$, with the metric and tensor normalization stated above. The retained coefficients are
\begin{equation}\label{14}
\begin{aligned}
\mathbbm w=(\mathbbm s,\mathbbm v_i,\mathbbm a_i,\mathbbm t_i),
\qquad \mathbbm t_i=\mathbbm T_{0i}-\mathbbm T_{i0}.
\end{aligned}
\end{equation}
When the system is in its initial vacuum state, the non-zero terms of the Wigner function are:
\begin{equation}\label{15}
	\begin{aligned}
		{{\mathbbm{s}}_{vac}} = \frac{{-2m}}{{\sqrt {{{\bf{p}}^2} + {m^2}} }} , ~~~~~{{\mathbbm{v}}_{i,vac}}=\frac{-2{{p}_{i}}}{\sqrt{{{\mathbf{p}}^{2}}+{{m}^{2}}}}.
	\end{aligned}
\end{equation}
The phase-space energy density of the fermion field is
\begin{equation}\label{16}
	\begin{aligned}
		\varepsilon  = m\mathbbm{s} + {p_i}{{\mathbbm{v}}_{i}}.
	\end{aligned}
\end{equation}
The single-particle distribution function can be expressed as the difference between the energy density and vacuum energy density:
\begin{equation}\label{17}
	\begin{aligned}
		f({\bf{q}},t) = \frac{1}{{2{\rm{\Omega }}({\bf{q}},t)}}\left( {\varepsilon  - {\varepsilon _{vac}}} \right),
	\end{aligned}
\end{equation}
where $\Omega(\mathbf q,t)=\sqrt{m^2+\mathbf p^2(t)}$ is the instantaneous one-particle energy and $\varepsilon_{\mathrm{vac}}=-2\Omega$. In this normalization, $f$ is the spin-summed electron occupation; its asymptotic value counts one electron per created pair. A particle interpretation during the pulse depends on the instantaneous basis, whereas the final spectrum is defined in the asymptotic free basis.

For numerical evolution, introduce the auxiliary vector $\mathbf v=\mathbbm{v}-(1-f)\mathbbm{v}_{\mathrm{vac}}$. Then $\mathbbm s=(1-f)\mathbbm s_{\mathrm{vac}}-\mathbf p\cdot\mathbf v/m$. The axial and tensor vectors are unchanged. Equation~\eqref{13} becomes~\cite{amat2023effects,majczak2024scattering,aleksandrov2024kinetic,amat2025spin,zdf5-w21r}
\begin{equation}\label{20}
\begin{aligned}
\dot f&=\frac{e}{2\Omega}\mathbf E\cdot\mathbf v,\\[1mm]
\dot{\mathbf v}&=\frac{2}{\Omega^3}
\bigl[(e\mathbf E\cdot\mathbf p)\mathbf p-e\Omega^2\mathbf E\bigr](f-1)\\
&\quad-\frac{(e\mathbf E\cdot\mathbf v)\mathbf p}{\Omega^2}
-2\mathbf p\times\mathbbm{a}-2m\mathbbm{t},\\[1mm]
\dot{\mathbbm{a}}&=-2\mathbf p\times\mathbf v,\\[1mm]
\dot{\mathbbm{t}}&=\frac2m
\bigl[m^2\mathbf v+(\mathbf p\cdot\mathbf v)\mathbf p\bigr].
\end{aligned}
\end{equation}
The vacuum initial conditions are $f(\mathbf q,-\infty)=0$ and $\mathbf v(\mathbf q,-\infty)=\mathbbm{a}(\mathbf q,-\infty)=\mathbbm{t}(\mathbf q,-\infty)=\mathbf0$. Dots denote time derivatives along the characteristic at fixed $\mathbf q$. The electric field and vector potential satisfy $\dot{\mathbf p}=e\mathbf E$ and $\mathbf E=-\dot{\mathbf A}$.

The pair number density is obtained by integrating the asymptotic spin-summed electron distribution over all three momentum components~\cite{kohlfurst2018phase,hebenstreit2010relating}.
\begin{equation}\label{21}
	\begin{aligned}
		n=\underset{t\to +\infty }{\mathop{\lim }}\,\int \frac{{{d}^{3}}q}{{{(2\pi )}^{3}}}f(\mathbf{q},t).
	\end{aligned}
\end{equation}
To define spin-resolved electron spectra, we use the asymptotic free-particle spin basis obtained by boosting rest-frame Pauli spinors, as specified in Appendix~\ref{APPENDIX}. Let $\mathbf n$ be a unit vector in that rest-frame spin space. This Foldy--Wouthuysen spin prescription is a defined observable; it is not a direct projection of the unboosted Dirac spin matrix along an arbitrary laboratory axis. Its spin-resolved energy combinations are
\begin{widetext}
\begin{equation}\label{22}
\begin{aligned}
\varepsilon_{\mathbf n,\pm}&=\frac12
\bigl(m\mathbbm s+\mathbf p\cdot\mathbbm{v}\bigr)\pm\frac12\biggl[m\mathbf n\cdot\mathbbm{a}+\mathbf n\cdot(\mathbf p\times\mathbbm{t})+\frac{(\mathbf n\cdot\mathbf p)
(\mathbf p\cdot\mathbbm{a})}{\Omega+m}\biggr].
\end{aligned}
\end{equation}
\end{widetext}
Vacuum subtraction gives the two spin-resolved distributions
\begin{equation}\label{23}
\begin{aligned}
f_{\mathbf n,\pm}(\mathbf q,t)
=\frac12\bigl[f(\mathbf q,t)\pm\delta f_{\mathbf n}(\mathbf q,t)\bigr].
\end{aligned}
\end{equation}
In the specified basis, the spin difference is related to the raw equal-time axial and tensor components by Eq.~\eqref{A29}~\cite{aleksandrov2024kinetic,amat2025spin}:
\begin{widetext}
\begin{equation}\label{24}
\begin{aligned}
\delta f_{\mathbf n}&=\frac1{2\Omega}\biggl[
 m\mathbf n\cdot\mathbbm{a}
 +\mathbf n\cdot(\mathbf p\times\mathbbm{t})+\frac{(\mathbf n\cdot\mathbf p)
(\mathbf p\cdot\mathbbm{a})}{\Omega+m}\biggr]=-2\mathbf n\cdot\mathbf F.
\end{aligned}
\end{equation}
\end{widetext}
For the figure comparisons, the intended convention is $\mathbf n=\hat{\mathbf z}$, with $f_{\mathrm{up}}=f_{\hat{\mathbf z},+}$ and $f_{\mathrm{down}}=f_{\hat{\mathbf z},-}$. On $p_z=q_z=0$, Eq.~\eqref{24} reduces to $[m\mathbbm a_z+(\mathbf p\times\mathbbm{t})_z]/(2\Omega)$. The numerical implementation must use this same projector; other DHW spin prescriptions can differ even on $q_z=0$ when their normalization or axis definition differs. The integrated spin-resolved number densities are
\begin{equation}\label{25}
\begin{aligned}
n_{\mathbf n,\pm}=\lim_{t\to+\infty}
\int\frac{d^3q}{(2\pi)^3}\,f_{\mathbf n,\pm}(\mathbf q,t).
\end{aligned}
\end{equation}

\section{Numerical results and discussion}\label{Discussion}
Vacuum pair production can be dominated by tunneling or by multiphoton absorption, with intermediate regimes possible. We use the reference Keldysh parameter~\cite{keldysh1965ionization}
\begin{equation}\label{26}
\begin{aligned}
\gamma_0=\frac{m\omega}{|e|E_0},
\end{aligned}
\end{equation}

where the peak-field parameter is $\gamma_{\mathrm{pk}}=\gamma_0\sqrt{1+\delta^2}$. When $\gamma_{\mathrm{pk}} \ll 1$ , the tunneling process dominates pair production and it is called nonperturbative Schwinger pair production. When $\gamma_{\mathrm{pk}} \gg 1$, the main contribution to pair creation comes from the multiphoton mechanism and this process is called perturbative multiphoton pair creation. For a chirped finite pulse, these parameters classify reference regimes rather than the whole evolving spectrum. Values much smaller than unity suggest predominantly tunneling dynamics, while values much larger than unity suggest multiphoton dynamics.

\subsection{High-frequency electric fields}\label{high}
For the high-frequency case, $E_0=0.1E_{\mathrm{cr}}$, $\omega=0.9m$, and $\tau=100/m$. Thus $\gamma_0=9$ and $\gamma_{\mathrm{pk}}\in[9,9\sqrt2]$. We use electron occupations in the final state convention defined above. Interpreting the supplied channels in this convention requires checking the charge and spin labels in the numerical implementation. The plot label $n$ denotes $f$ on a momentum plane, rather than the integrated density in Eq.~\eqref{21}.
\begin{figure*}
	\centering
	\includegraphics[width=0.8\linewidth]{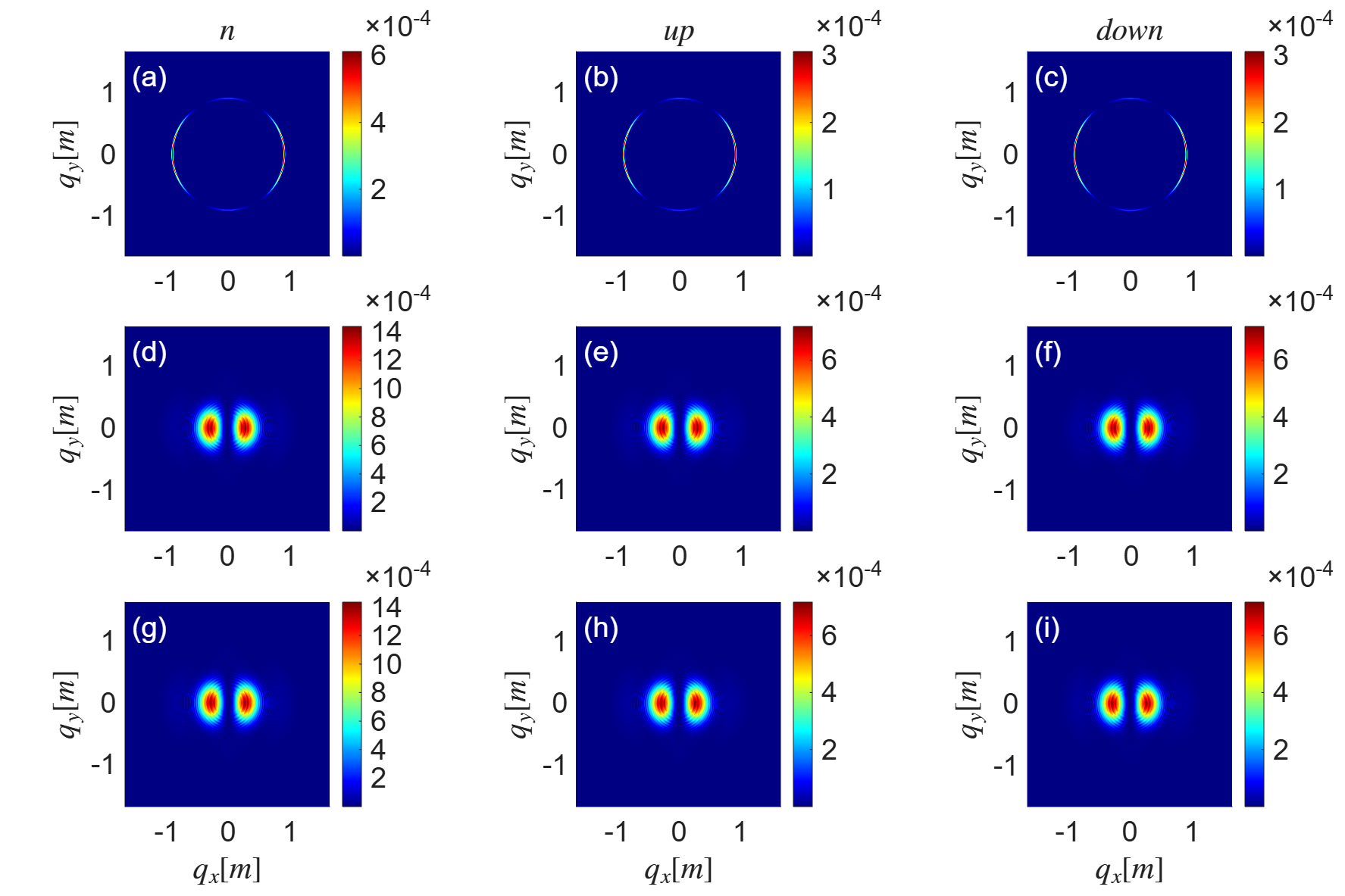}
	\caption{Asymptotic electron momentum spectra on the $(q_x,q_y)$ plane at $q_z=0$ for linear polarization ($\delta=0$), with $\omega=0.9m$, $E_0=0.1E_{\mathrm{cr}}$, and $\tau=100/m$. Columns show the spin-summed distribution $f$ (labeled $n$), $f_{\mathrm{up}}$, and $f_{\mathrm{down}}$ in the basis of Eq.~\eqref{24}. Rows correspond to $\beta=b\tau/\omega=0$, $0.1$, and $-0.1$. Independent color bars indicate the scale of each channel.}
	\label{fig1}
\end{figure*}
\begin{figure*}
	\centering
	\includegraphics[width=0.8\linewidth]{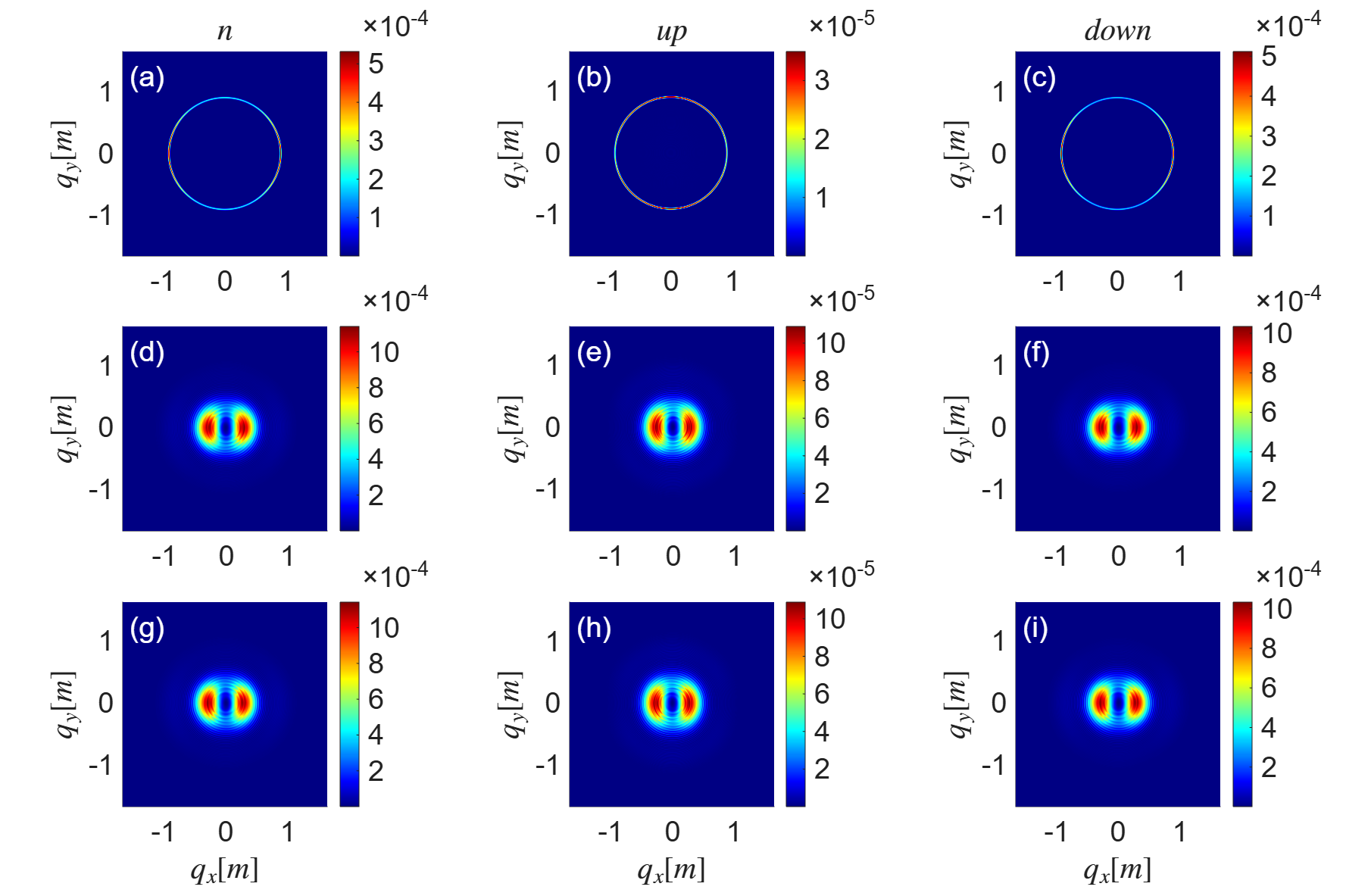}
	\caption{Asymptotic electron momentum spectra on the $(q_x,q_y)$ plane at $q_z=0$ for elliptical polarization ($\delta=0.5$), with $\omega=0.9m$, $E_0=0.1E_{\mathrm{cr}}$, and $\tau=100/m$. Columns show the spin-summed distribution $f$ (labeled $n$), $f_{\mathrm{up}}$, and $f_{\mathrm{down}}$ in the basis of Eq.~\eqref{24}. Rows correspond to $\beta=b\tau/\omega=0$, $0.1$, and $-0.1$. Independent color bars indicate the scale of each channel.}
	\label{fig2}
\end{figure*}

\begin{figure*}
	\centering
	\includegraphics[width=0.8\linewidth]{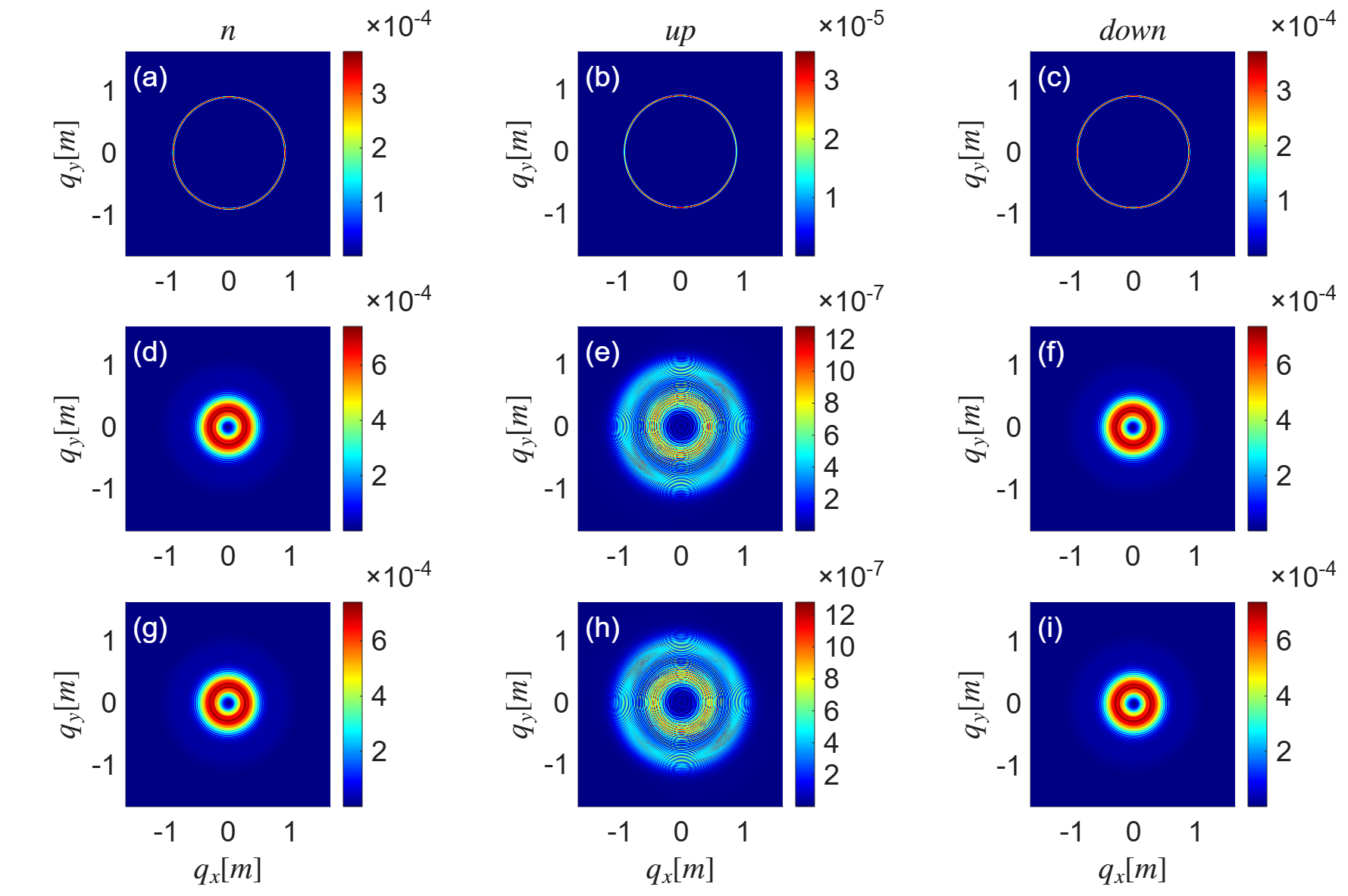}
	\caption{Asymptotic electron momentum spectra on the $(q_x,q_y)$ plane at $q_z=0$ for circular polarization ($\delta=1$), with $\omega=0.9m$, $E_0=0.1E_{\mathrm{cr}}$, and $\tau=100/m$. Columns show the spin-summed distribution $f$ (labeled $n$), $f_{\mathrm{up}}$, and $f_{\mathrm{down}}$ in the basis of Eq.~\eqref{24}. Rows correspond to $\beta=b\tau/\omega=0$, $0.1$, and $-0.1$. Independent color bars indicate the scale of each channel.}
	\label{fig3}
\end{figure*}

In this prescribed background, $\mathbf B=0$, so a magnetic Zeeman splitting is absent. Spin asymmetry instead reflects the time-dependent Dirac evolution and the unequal transition probabilities in a chosen spin basis~\cite{strobel2015,blinne2016,amat2025spin}. The field direction enters the spin-basis rotation and positive--negative-energy couplings in Appendix~\ref{APPENDIX}. This does not establish an axial-anomaly mechanism: the electromagnetic anomaly source proportional to $\mathbf E\cdot\mathbf B$ vanishes here, although massive axial dynamics can still occur~\cite{adler1969axial,bell1969pcac,copinger2018axial}. Ellipticity is a field-shape parameter, not an added axial vector potential. Likewise, a spin-current interpretation would require calculating a defined current observable~\cite{huang2019spincurrent}.

We first consider linear polarization ($\delta=0$). Fig.~\ref{fig1} shows electron spectra on the $(q_x,q_y)$ plane at $q_z=0$. Its columns give the spin-summed and two spin-resolved distributions, while its rows correspond to $\beta=0$, $0.1$, and $-0.1$. This is a comparison of momentum-dependent occupations, not three integrated number densities.

For $b=0$, Fig.~\ref{fig1}(a) exhibits a narrow resonance ring with its largest values near the polarization axis. This is consistent with a multiphoton-dominated spectrum. In a long, weak-field pulse, the leading resonance estimate is $N\omega\simeq2\sqrt{m_*^2+\mathbf q^2}$, where $m_*$ accounts for field dressing~\cite{kohlfurst2014effective}. Finite duration produces a resonance width, so the relation is an approximation rather than an exact energy-conservation condition for this pulse.

Figs.~\ref{fig1}(b) and \ref{fig1}(c) have identical shapes and peak occupations, approximately $3\times10^{-4}$. Their equality is consistent with the spin-degenerate sector of the fixed-direction electric-field problem in the chosen basis. In a photon description, linear polarization has balanced helicity components, but this observation alone is not a derivation of momentum-by-momentum spin equality. Chirping a field of fixed direction preserves this spin degeneracy while modifying the common spectral profile.

For elliptical polarization ($\delta=0.5$), Fig.~\ref{fig2} shows how a rotating field changes the two spin channels. The reference parameter remains $\gamma_0=9$, while the actual peak-field parameter is $\gamma_{\mathrm{pk}}=9\sqrt{1.25}\simeq10.1$. The field is therefore in the high-frequency reference regime. Its time-dependent direction allows spin-selective transitions that are absent in the fixed-direction case; the sign of the favored channel depends on the handedness, charge, and spin-label conventions.

In the unchirped spectra, Figs.~\ref{fig2}(a)--\ref{fig2}(c), the two spin populations differ while both retain a resonance-ring structure. This is consistent with a dominant three-quantum channel because $3\omega>2m$. A rotating background selects an oriented field history and can distinguish the spin channels; this should not be described as a violation of parity symmetry by QED. Its spatial and temporal symmetry properties must instead be evaluated for the specified external field.

For the unchirped carrier $\omega=0.9m$, one and two carrier quanta lie below the free-pair threshold, while $3\omega=2.7m>2m$. Neglecting the small dressing correction gives $3\omega\simeq2\sqrt{m^2+q_x^2+q_y^2}$ at $q_z=0$, and hence a resonance radius $|\mathbf q_\perp|\simeq0.907m$. This estimate explains the location of the narrow ring but does not require an exactly circular or infinitely sharp spectrum: the envelope, field dressing, angular transition matrix elements, and interference remain relevant. With chirp, the phase frequency sweeps as $\omega+2bt$ and the spectral bandwidth increases. In parts of a positive-frequency sweep where $\omega_{\mathrm{eff}}>m$, even a two-quantum threshold can be crossed. The broadened spectra therefore cannot be assigned universally to one fixed three-photon channel without a channel-resolved calculation.

The color bars in Fig.~\ref{fig2} indicate peak occupations of approximately $5\times10^{-4}$ in the displayed spin-down channel and $3.5\times10^{-5}$ in the spin-up channel. The difference demonstrates spin-selective production in this basis, and the dominant channel largely determines the spin-summed profile. These are maxima on the plotted momentum window; they are not integrated yields or direct measurements of angular-momentum transfer. A full angular-momentum balance would also require the orbital contribution and a treatment of the background.

\begin{figure*}
	\centering
	\includegraphics[width=0.9\linewidth]{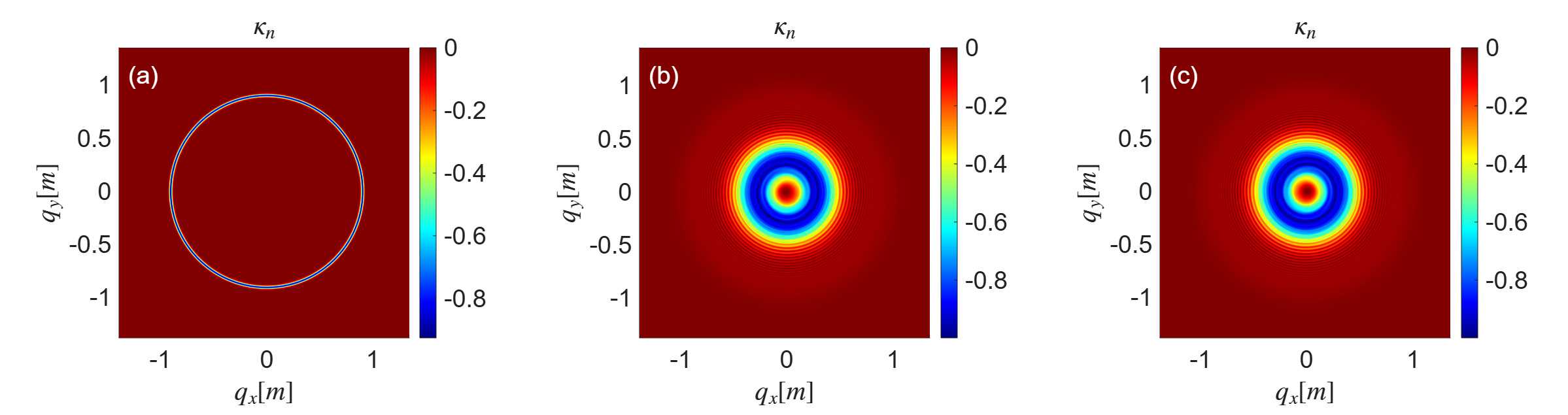}
	\caption{Normalized spin-difference spectrum $\kappa_n=(f_{\mathrm{up}}-f_{\mathrm{down}})/f_{\max}$ on the displayed $(q_x,q_y)$ window at $q_z=0$ for circular polarization ($\delta=1$), with $\omega=0.9m$, $E_0=0.1E_{\mathrm{cr}}$, and $\tau=100/m$. Panels (a)--(c) have $\beta=b\tau/\omega=0$, $0.1$, and $-0.1$, respectively. Here $f_{\max}$ is the maximum spin-summed spectrum on the same window for each configuration; $\kappa_n$ is not the local polarization.}
	\label{fig4}
\end{figure*}

At $\delta=1$, the field magnitude is circularly polarized within a Gaussian envelope, as shown in Fig.~\ref{fig3}. The unchirped spectra have an approximately circular resonance ring. For a long rotating pulse, angular averaging can make the dominant spectrum nearly isotropic. A finite pulse with a specified carrier phase does not, however, have an exact continuous rotational symmetry at all times; the weak channel can retain angular structure.

Spin selectivity is especially pronounced in the chirped circular spectra. In Figs.~\ref{fig3}(f) and \ref{fig3}(i), the displayed spin-down peak is about $7\times10^{-4}$, compared with about $1.2\times10^{-6}$ in Figs.~\ref{fig3}(e) and \ref{fig3}(h), a difference of several hundred. The spin axis remains fixed at $\mathbf n=\hat{\mathbf z}$ throughout this comparison. The normalized difference $\kappa_n$ is negative where the spin-down channel dominates and approaches $-1$ near a sufficiently dominant peak. Since $\kappa_n=P_z f/f_{\max}$, a value near $-1$ implies strong local polarization there, but a small value away from the spectral maximum can also reflect a small total occupation. Integrated polarization cannot be inferred from this plot alone.

Chirping changes the narrow ring into a broader radial band with interference structure, as seen in Fig.~\ref{fig3}. This broadening reflects the changed frequency content and coherent field history. It does not by itself prove an increase in the full momentum-integrated density, which requires the integral in Eq.~\eqref{21}. A temporal-interference interpretation is plausible, but identifying exactly two symmetric creation times would require a turning-point or transition-amplitude analysis. The chirped phase $\omega t+bt^2$ does not generally give symmetric resonance events at $\pm t_0$.

Fig.~\ref{fig4} displays $\kappa_n$ for circular polarization. This observable locates the spin-channel difference relative to the largest spin-summed occupation in each plotted window. It therefore combines spin imbalance with spectral weight. The linear-polarization limit has $\kappa_n=0$ in the selected basis, whereas the circular case produces a predominantly negative structure. We use this spectral descriptor to compare the pulse configurations, while retaining the separate definitions of local and integrated polarization.

In the unchirped case, Fig.~\ref{fig4}(a), the normalized difference is concentrated on the resonance ring. Its negative sign is consistent with the larger displayed spin-down population. The strength and angular variation must be read together with the individual channel color scales in Fig.~\ref{fig3}; peak values alone do not establish uniform spin polarization around the whole ring. The relation $f_{\mathrm{up}}+f_{\mathrm{down}}=f$ must hold pointwise before a quantitative comparison of these plots is made.

For $\beta=\pm0.1$, Figs.~\ref{fig4}(b) and \ref{fig4}(c), the negative spin-difference structure broadens and develops radial modulation. The near equality of the positive- and negative-chirp high-frequency patterns is visible in the displayed spectra. These features are consistent with coherent production over a broadened range of frequencies. They do not establish isotropic phase accumulation at every momentum or a uniform maximum of $|\kappa_n|$ over the entire band. Claims about a macroscopic polarized source would require integrated spectra, absolute rates, and an experimentally realizable field configuration.

\begin{figure}
	\centering
	\includegraphics[width=0.9\linewidth]{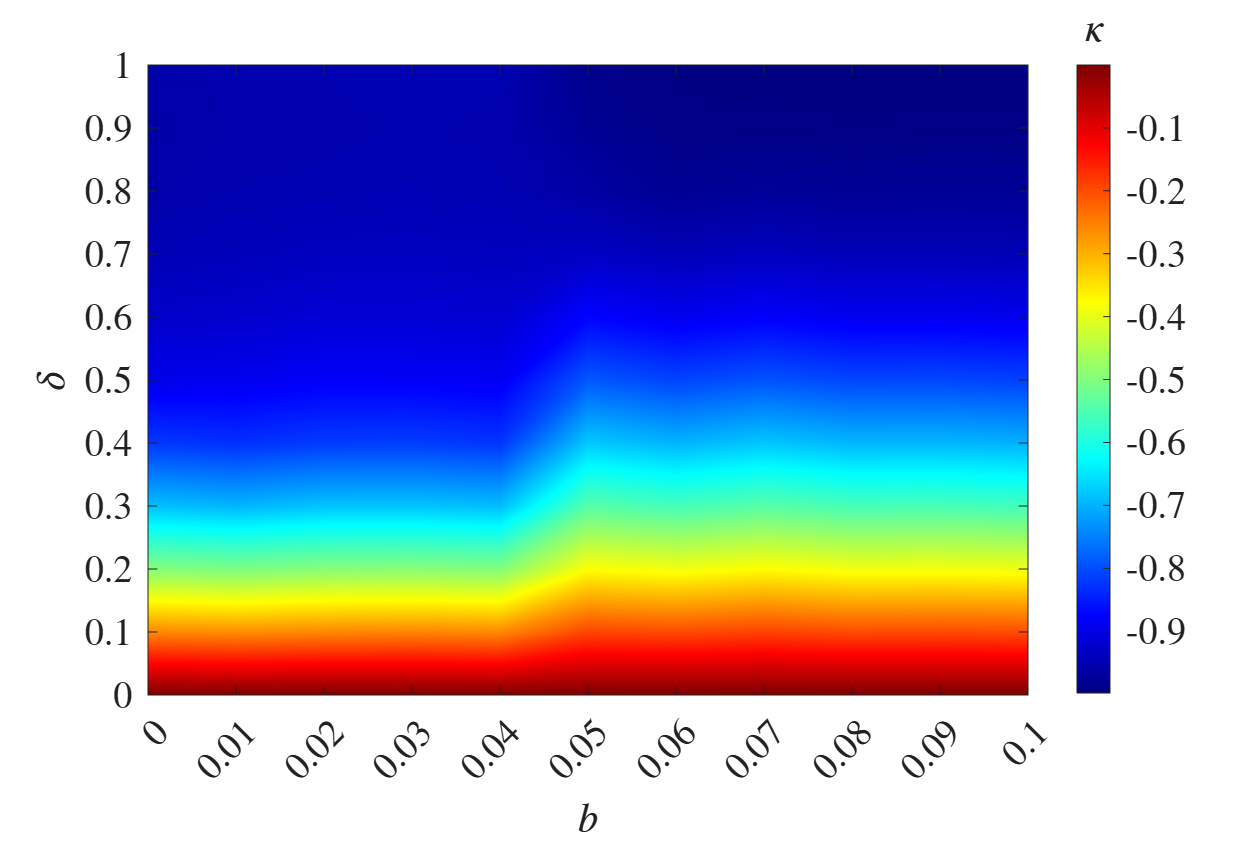}
	\caption{Peak-asymmetry indicator $\kappa=(f_{\max}^{\mathrm{up}}-f_{\max}^{\mathrm{down}})/f_{\max}$ versus $\delta$ and $\beta=b\tau/\omega$ (the horizontal axis is labeled $b$), at $\omega=0.9m$, $E_0=0.1E_{\mathrm{cr}}$, and $\tau=100/m$. All maxima refer to the displayed $q_z=0$ momentum window. The two channel maxima need not be at the same momentum; this indicator is distinct from $\kappa_n$ and from integrated polarization.}
	\label{fig5}
\end{figure}

\begin{figure*}
	\centering
	\includegraphics[width=0.9\linewidth]{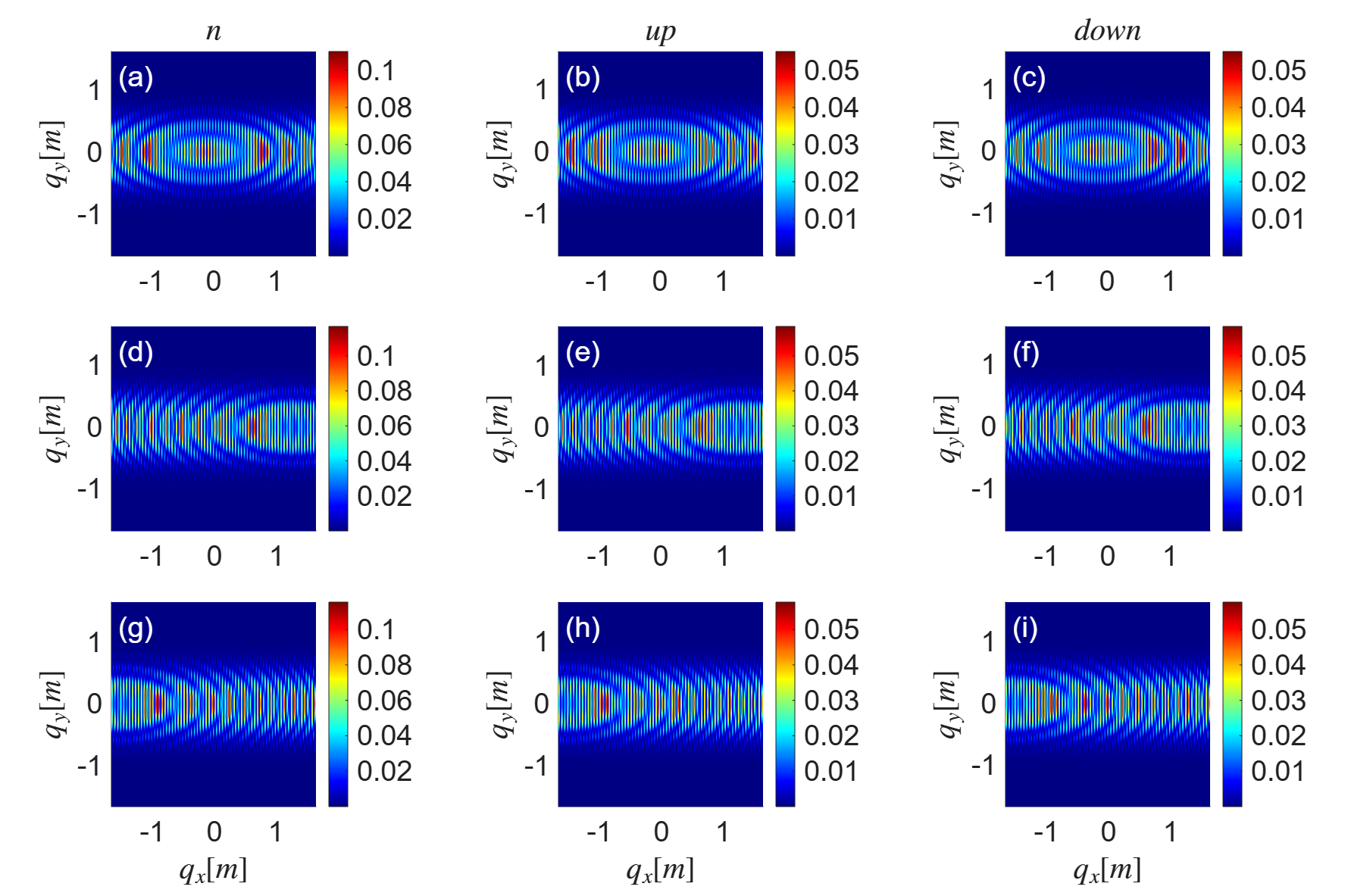}
	\caption{Asymptotic electron momentum spectra on the $(q_x,q_y)$ plane at $q_z=0$ for linear polarization ($\delta=0$), with $\omega=0.05m$, $E_0=0.5\sqrt2E_{\mathrm{cr}}$, and $\tau=100/m$. Columns show the spin-summed distribution $f$ (labeled $n$), $f_{\mathrm{up}}$, and $f_{\mathrm{down}}$ in the basis of Eq.~\eqref{24}. Rows correspond to $\beta=b\tau/\omega=0$, $0.1$, and $-0.1$. Independent color bars indicate the scale of each channel.}
	\label{fig6}
\end{figure*}

Fig.~\ref{fig5} shows the peak indicator $\kappa$ as a function of ellipticity and $\beta=b\tau/\omega$ at $\omega=0.9m$. Its numerator subtracts two independently located maxima. Consequently, this map summarizes a spectral contrast and is distinct from a map of either $P_z$ or $\mathcal P_z$.

The ellipticity dependence shows $\kappa\simeq0$ near linear polarization and increasingly negative values as the displayed spin-down channel becomes dominant. This trend is consistent with stronger spin selectivity in rotating fields. The color map alone does not establish an exponential law for suppression of the other channel. Since the normalization changes the actual peak field with $\delta$, the scan also combines changes in field direction with changes in peak amplitude.

The chirp dependence modifies the ellipticity at which large negative peak asymmetry appears. The plotted contour changes most visibly near $\beta\simeq0.04$--$0.05$, after which comparable contrast extends to somewhat smaller $\delta$. This is a smooth crossover in the displayed parameter scan, rather than an established critical point or phase transition. Precise threshold values would require the underlying grid data and a declared numerical contour criterion.

The expanded region of large negative $\kappa$ suggests that frequency chirping can alter spin-selective multiphoton production for this pulse family~\cite{olugh2019pair,li2021chirp}. Broadening of the frequency content and changed temporal interference are possible contributions. Separating these effects requires additional controls, such as comparing spectra with matched bandwidth and peak amplitude. The scan does not by itself demonstrate a specific angular-momentum absorption pathway or a dynamically assisted two-field mechanism.

\subsection{Low-frequency electric fields}\label{low}
\begin{figure*}
	\centering
	\includegraphics[width=0.9\linewidth]{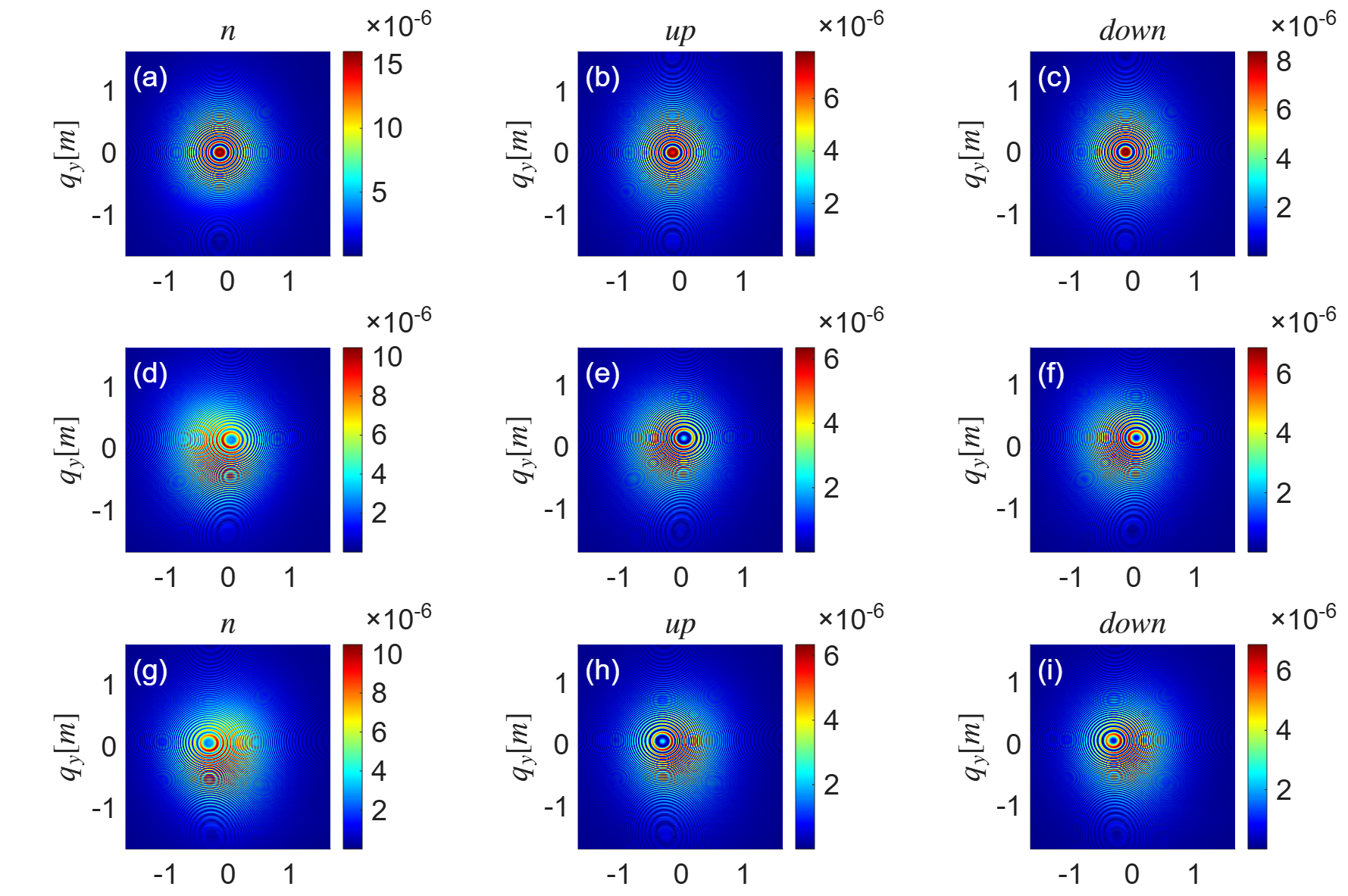}
	\caption{Asymptotic electron momentum spectra on the $(q_x,q_y)$ plane at $q_z=0$ for elliptical polarization ($\delta=0.5$), with $\omega=0.05m$, $E_0=0.5\sqrt2E_{\mathrm{cr}}$, and $\tau=100/m$. Columns show the spin-summed distribution $f$ (labeled $n$), $f_{\mathrm{up}}$, and $f_{\mathrm{down}}$ in the basis of Eq.~\eqref{24}. Rows correspond to $\beta=b\tau/\omega=0$, $0.1$, and $-0.1$. Independent color bars indicate the scale of each channel.}
	\label{fig7}
\end{figure*}

\begin{figure*}
	\centering
	\includegraphics[width=0.9\linewidth]{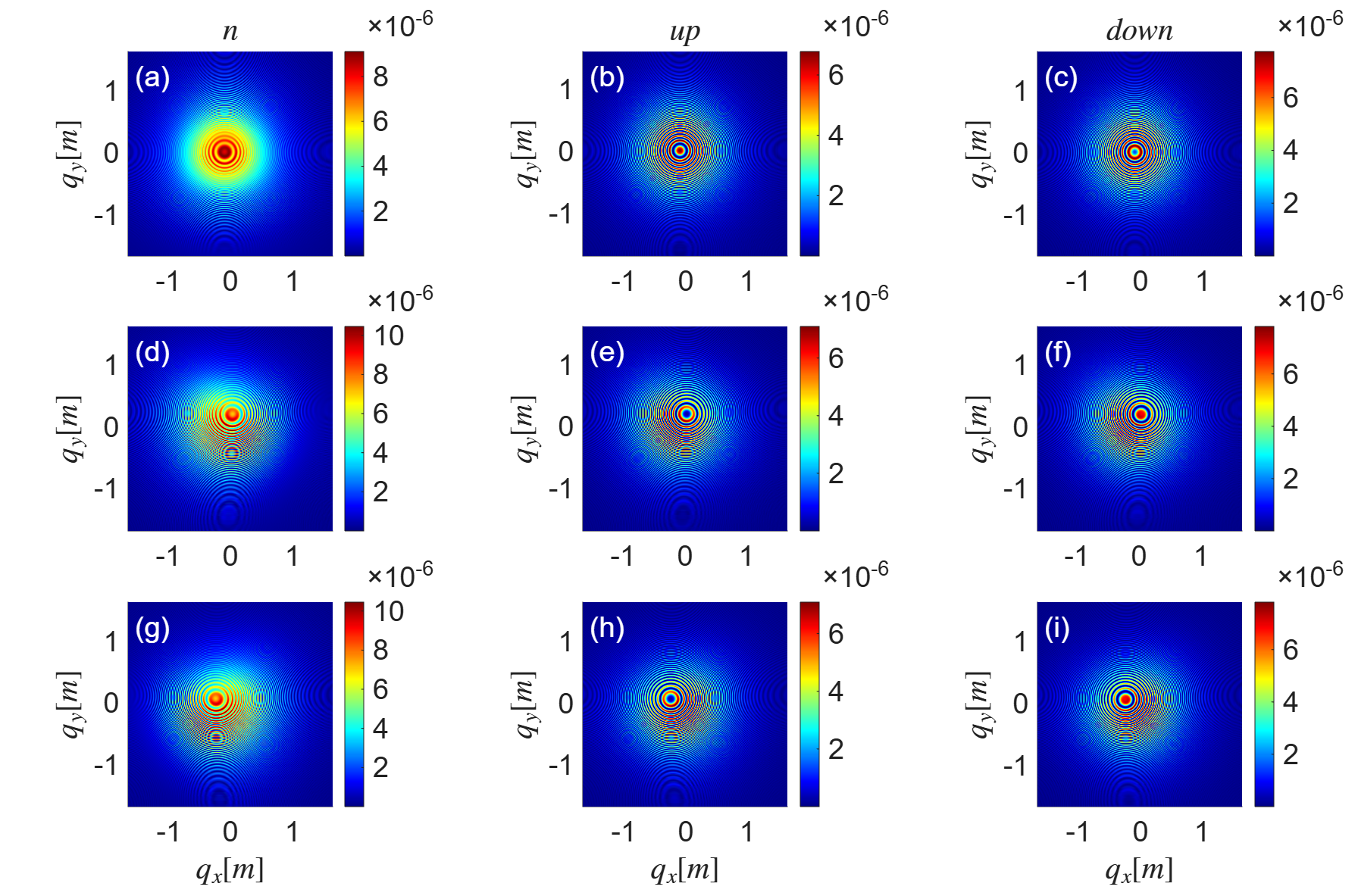}
	\caption{Asymptotic electron momentum spectra on the $(q_x,q_y)$ plane at $q_z=0$ for circular polarization ($\delta=1$), with $\omega=0.05m$, $E_0=0.5\sqrt2E_{\mathrm{cr}}$, and $\tau=100/m$. Columns show the spin-summed distribution $f$ (labeled $n$), $f_{\mathrm{up}}$, and $f_{\mathrm{down}}$ in the basis of Eq.~\eqref{24}. Rows correspond to $\beta=b\tau/\omega=0$, $0.1$, and $-0.1$. Independent color bars indicate the scale of each channel.}
	\label{fig8}
\end{figure*}

\begin{figure*}
	\centering
	\includegraphics[width=0.9\linewidth]{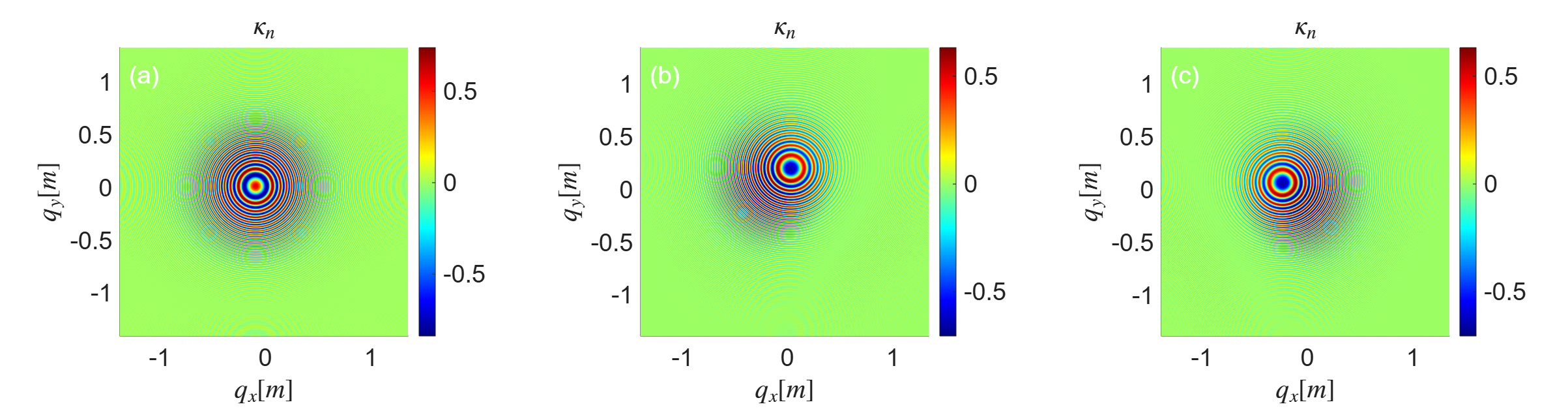}
	\caption{Normalized spin-difference spectrum $\kappa_n=(f_{\mathrm{up}}-f_{\mathrm{down}})/f_{\max}$ on the displayed $(q_x,q_y)$ window at $q_z=0$ for circular polarization ($\delta=1$), with $\omega=0.05m$, $E_0=0.5\sqrt2E_{\mathrm{cr}}$, and $\tau=100/m$. Panels (a)--(c) have $\beta=b\tau/\omega=0$, $0.1$, and $-0.1$, respectively. Here $f_{\max}$ is the maximum spin-summed spectrum on the same window for each configuration; $\kappa_n$ is not the local polarization.}
	\label{fig9}
\end{figure*}

For the low-frequency case, $E_0=0.5\sqrt2E_{\mathrm{cr}}$, $\omega=0.05m$, and $\tau=100/m$. Thus $\gamma_0\simeq0.0707$ and $\gamma_{\mathrm{pk}}$ ranges from approximately $0.0707$ to $0.1$ as $\delta$ varies from $0$ to $1$. These values identify an adiabatic reference regime in which tunneling is expected to dominate. Finite duration and rotation can nevertheless produce temporal interference and spin-dependent corrections. The plotted momentum window need not contain the entire low-frequency spectrum, so a comparison of its peaks is not a comparison of total yields.

For linear polarization, Fig.~\ref{fig6}, the two spin-resolved spectra are identical in shape and have peaks near $5\times10^{-2}$. Equality of the two channels in the fixed-direction problem implies zero spin difference in the chosen basis, consistently in both frequency regimes. The visible modulation reflects coherent creation and subsequent evolution over the pulse; the number of fringes does not provide an independent spin observable.

The dominant structures in Fig.~\ref{fig6} are interference fringes approximately parallel to the $q_y$ axis in the unchirped case. Successive field extrema can act as separate creation intervals~\cite{dumlu2010stokes,hebenstreit2009momentum}. The relative dynamical phase between positive--negative-energy transition amplitudes is estimated by~\cite{dumlu2011interference}
\begin{equation}
\begin{aligned}
\Delta\Phi_{\mathrm{dyn}}(\mathbf q)&=2\int_{t_k}^{t_{k+1}}\Omega(\mathbf q,t)\,dt,\\
\Omega(\mathbf q,t)&=\sqrt{m^2+[q_x-eA_x(t)]^2+q_y^2},
\end{aligned}
\end{equation}
at $q_z=0$. The factor $2$ is the energy gap between the two branches. Additional connection and turning-point phases can enter the interference condition. In the central part of a slowly modulated unchirped pulse, adjacent extrema are separated by approximately $\pi/\omega$; this is not exact throughout a Gaussian envelope. The observed fringe orientation is consistent with the stronger longitudinal momentum dependence, while $q_y$ enters through the transverse energy.

For $\beta=\pm0.1$, Figs.~\ref{fig6}(d)--\ref{fig6}(i), the instantaneous phase frequency $\omega_{\mathrm{eff}}=\omega+2bt$ changes the timing and phases of successive creation intervals. The fringes become curved and their envelope shifts. In the supplied figure, positive chirp places the more prominent structure toward positive $q_x$, while negative chirp shifts it toward negative $q_x$. This comparison fixes the descriptive direction from the displayed spectra; it should not be interpreted as a gauge-independent momentum kick until the vector-potential boundary convention is specified.

For elliptical polarization ($\delta=0.5$), Fig.~\ref{fig7} displays closed interference contours near the origin together with angular structure. At $b=0$, the central contours are approximately concentric, but the full spectrum is not exactly isotropic. Although $|\mathbf E(t)|$ reaches its global maximum at $t=0$, $\mathbf A(0)$ need not vanish; in particular, the rotating field generally has a nonzero $A_y(0)$ in the final-zero gauge. A central interference pattern therefore cannot be attributed simply to tunneling at zero canonical momentum at the field maximum.

A finite chirp changes the vector potential and the relative weights and phases of creation amplitudes~\cite{olugh2019pair,osman2023efficient,li2021chirp}. It does not shift the global maximum of $|\mathbf E|$ for Eq.~\eqref{1}. If a single tunneling interval at $t_*$ dominates and the initial kinetic momentum is small, $\mathbf p(t_*)\simeq\mathbf0$ implies $\mathbf q\simeq e\mathbf A(t_*)$. The final kinetic momentum is then $e[\mathbf A(t_*)-\mathbf A(+\infty)]$, with the signed electron charge. This estimate cannot generally locate the center of a multicycle interference pattern. In Fig.~\ref{fig7}, positive and negative chirp change the central contours and their displacement in $q_x$; the detailed position reflects the full coherent evolution, rather than an assumed shift of one field maximum.

The unchirped channel peaks in Fig.~\ref{fig7} are comparable: approximately $8\times10^{-6}$ for the displayed spin-down channel and $7\times10^{-6}$ for the spin-up channel. This gives a smaller contrast between maxima than in the high-frequency examples. It does not imply a comparably small local spin difference at every momentum, since the two channels can have shifted interference fringes. The leading locally constant electric-field tunneling exponent is spin independent, but finite-pulse and rotating-field corrections can remain spin dependent. Angular-momentum conservation remains valid in the tunneling regime.

For circular polarization ($\delta=1$), Fig.~\ref{fig8} again shows comparable spin-channel peaks of order $10^{-6}$. The unchirped central contours are approximately concentric, with visible angular structure farther from the center. The Gaussian envelope and fixed phase prevent an assumption of exact rotational symmetry for the full finite pulse. Similar peak magnitudes establish a small difference of maxima within the displayed window, while allowing a sizeable pointwise difference between the two oscillatory channels.

For $\beta=\pm0.1$, Figs.~\ref{fig8}(d)--\ref{fig8}(i), the central contours shift and deform. The phase accumulated over a short time interval satisfies $\Delta\phi\simeq\omega_{\mathrm{eff}}(t_*)\Delta t$, but this kinematic relation does not show that the field maximum or the dominant creation time has moved. Indeed, for $\delta=1$, $|\mathbf E(t)|=E_0\exp[-t^2/(2\tau^2)]/\sqrt2$ is independent of chirp. Reversing chirp changes the time history; in the displayed figure, the contours are approximately reflected in $q_x$ and do not move to exactly opposite quadrants.

\begin{figure}
	\centering
	\includegraphics[width=1\linewidth]{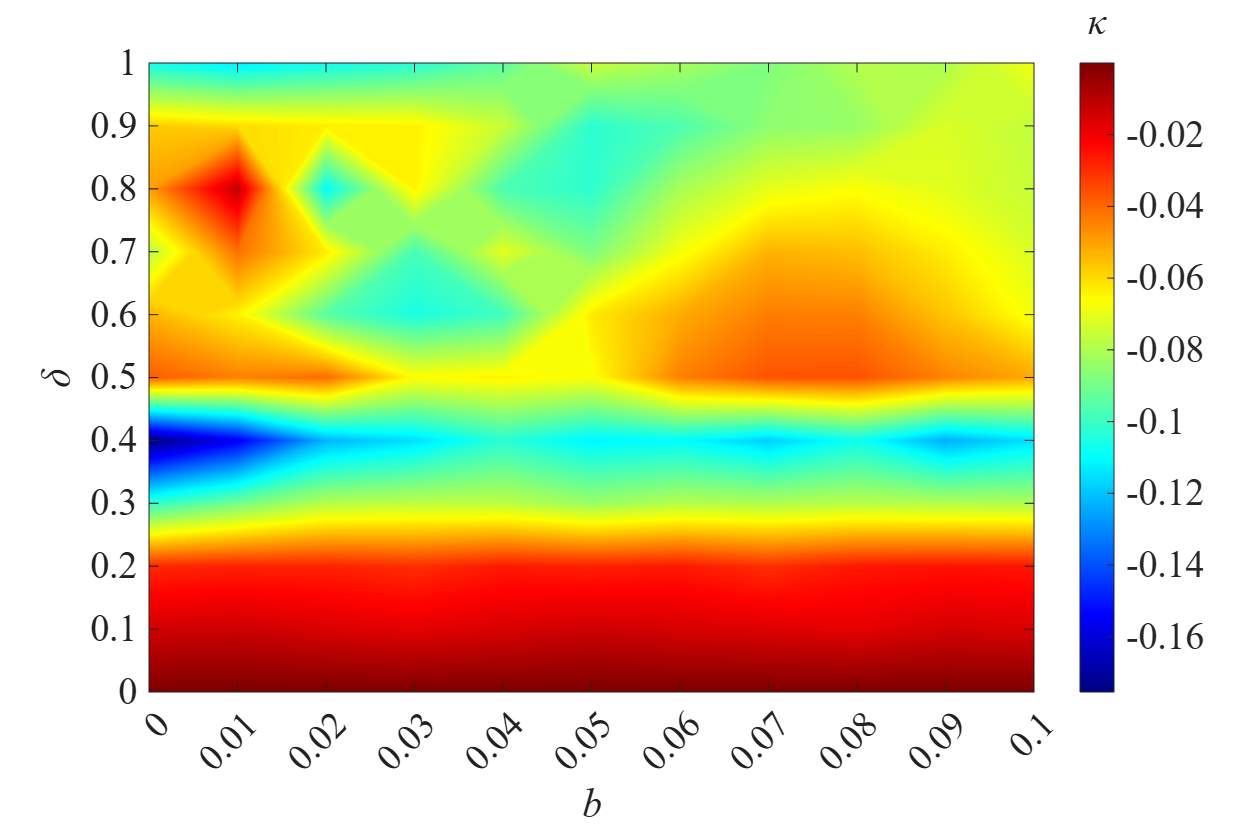}
	\caption{Peak-asymmetry indicator $\kappa=(f_{\max}^{\mathrm{up}}-f_{\max}^{\mathrm{down}})/f_{\max}$ versus $\delta$ and $\beta=b\tau/\omega$ (the horizontal axis is labeled $b$), at $\omega=0.05m$, $E_0=0.5\sqrt2E_{\mathrm{cr}}$, and $\tau=100/m$. All maxima refer to the displayed $q_z=0$ momentum window. The two channel maxima need not be at the same momentum; this indicator is distinct from $\kappa_n$ and from integrated polarization.}
	\label{fig10}
\end{figure}

Fig.~\ref{fig9} shows the normalized spin-difference spectrum $\kappa_n$ in the low-frequency circular case. Its sign-changing fringes differ markedly from the predominantly negative high-frequency pattern. The supplied color bars span values of order $\pm0.7$, so the claim that this spectrum is bounded by $0.2$ is not supported by the figure. The smaller value reported in Fig.~\ref{fig10} belongs to the difference-of-maxima indicator $\kappa$, which is a different quantity. The leading tunneling exponent $\exp[-\pi m^2/(|e|E)]$ is spin independent in a locally constant pure electric field, but it does not impose a small bound on spin-dependent finite-pulse interference.

The unchirped $\kappa_n$ spectrum in Fig.~\ref{fig9}(a) contains concentric alternating sign contours near the center. Taking $f_{\mathrm{up}}-f_{\mathrm{down}}$ emphasizes differences in the two coherent spectra and can reveal phase-sensitive structure even when their maxima are similar. Rotation of the spin basis is a possible contributor, as described by $\boldsymbol\mu_1$ in the QKEs. Establishing a purely phase-driven origin, however, requires comparing transition amplitudes and separating their phases from their magnitudes; the density plots alone do not provide that decomposition.

With $\beta=\pm0.1$, Figs.~\ref{fig9}(b) and \ref{fig9}(c), the alternating contours shift and deform along with the channel spectra. Their centers are approximately reflected in $q_x$ rather than exactly inverted about the origin. These changes demonstrate sensitivity to the pulse history. They do not establish that the pattern uniquely reconstructs the instantaneous phase or pulse waveform; such an inverse interpretation would require additional controls and a demonstrated reconstruction procedure.

Fig.~\ref{fig10} shows the low-frequency peak indicator $\kappa$ in the $(\delta,\beta)$ plane. It vanishes in the linear limit and remains within approximately $|\kappa|\lesssim0.18$ in the displayed scan. The smaller contrast between channel maxima is distinct from the larger sign-changing $\kappa_n$ fringes in Fig.~\ref{fig9}. The scan therefore supports a regime-dependent change in peak contrast, not a universal upper bound on spin polarization in tunneling. Discrete multiphoton selection rules are not directly applicable as a dominant-channel description here, while angular-momentum conservation still holds.

The chirp dependence in the low-frequency scan is nonmonotonic. Varying $\beta$ from $0$ to $0.1$ changes localized regions of peak contrast rather than producing the relatively smooth boundary displacement seen at high frequency. This behavior is consistent with competing coherent amplitudes in an adiabatic reference regime. Ellipticity also changes the actual peak field through Eq.~\eqref{1}, so its influence is not restricted to a subleading phase correction of a fixed tunneling exponent.

Overall, the low-frequency spectra exhibit small differences between channel maxima together with appreciable, oscillatory momentum-dependent spin differences. These observations should be stated separately. Determining the integrated asymmetry and its numerical uncertainty requires the full momentum spectra, a convergence analysis, and identical spin projectors in the DHW and QKT calculations.

\section{Conclusion} \label{sec:conclusion}
We have studied spin-resolved vacuum pair production in prescribed, spatially homogeneous electric fields with variable ellipticity and quadratic chirp, using the equal-time DHW formalism. The two parameter sets probe high-frequency multiphoton and low-frequency adiabatic reference regimes. Linearly polarized fields give identical distributions in the chosen spin basis, whereas rotating fields generate different spin-channel spectra. Chirping modifies both their momentum distributions and interference structure.

In the high-frequency circular case, the unchirped spectra contain a narrow resonance ring, while chirped spectra show a broader radial structure with pronounced dominance of the displayed spin-down channel. For the chirped examples, the two peak occupations differ by several hundred. In the low-frequency examples, their maxima are comparable, but their difference can exhibit substantial sign-changing fringes. A smaller contrast of maxima therefore does not imply that all local spin effects have been suppressed.

The parameter maps summarize the peak indicator $\kappa$. At high frequency, increasing ellipticity produces a large negative contrast, and chirp shifts the region where $|\kappa|\gtrsim0.9$. At low frequency, the reported scan gives $|\kappa|\lesssim0.18$ with nonmonotonic dependence on ellipticity and chirp. These bounds refer to the displayed momentum window and this specific definition. They are neither bounds on $\kappa_n$ nor measurements of the full momentum-integrated polarization.

The calculations show distinct spin-resolved spectral responses in the two reference regimes and identify chirp and ellipticity as controls of those responses. A physical interpretation must retain the spin projector, charge sign, momentum convention, and normalization. Quantitative statements about integrated polarization or experimental sources require full momentum integration, numerical error estimates, and realistic field modeling. The interference spectra motivate further study of spin-dependent coherent evolution, but a unique pulse-reconstruction method or an orbital-angular-momentum measurement has not been established here.

\begin{acknowledgments}
This work was supported by the National Natural Science Foundation of China (NSFC) (Grant No. 12564046). Orkash Amat was supported by the Specialized Research Fund for the State Key Laboratory of Radio Astronomy and Technology and by the China Postdoctoral Science Foundation (Grant No. 2025M783225).
\end{acknowledgments}

\appendix
\section{Verification by quantum kinetic equation} \label{APPENDIX}
We give the QKE for a prescribed, spatially homogeneous electric field of arbitrary time-dependent direction~\cite{aleksandrov2024kinetic,amat2025spin}. The derivation retains both intraband spin-basis rotation and interband mixing. It therefore does not use the approximation that neglects intraband connection terms in a simplified generalized TLM~\cite{amat2025spin}. We also give the exact transformation to the equal-time DHW variables~\cite{hebenstreit2010relating,bialynicki1991phasespace}. As in the main text, $\mathbf p(\mathbf q,t)=\mathbf q-e\mathbf A(t)$, $\Omega=\sqrt{m^2+\mathbf p^2}$, $A_0=0$, and $\mathbf E=-\dot{\mathbf A}$. The same signed charge, boundary condition, and final-state spin projector must be used in both descriptions.

For a homogeneous background, different canonical momenta do not mix. At fixed $\mathbf q$, the Dirac Hamiltonian is
\stepcounter{equation}
\begin{equation}\label{A1}
\begin{aligned}
H(\mathbf q,t)=\boldsymbol\alpha\cdot\mathbf p(\mathbf q,t)+\beta m,
\end{aligned}\tag{A1}
\end{equation}
Let $u_s(\mathbf p)$ and $v_s(\mathbf p)$ be orthonormal instantaneous eigenvectors with eigenvalues $+\Omega$ and $-\Omega$. Fix the basis as $u_s=N\binom{\chi_s}{(\boldsymbol\sigma\cdot\mathbf p)\chi_s/(\Omega+m)}$ and $v_s=N\binom{-(\boldsymbol\sigma\cdot\mathbf p)\chi_s/(\Omega+m)}{\chi_s}$, where $N=\sqrt{(\Omega+m)/(2\Omega)}$ and $\sigma_z\chi_s=s\chi_s$ for $s=\pm1$. This explicitly specifies the rest-frame spin convention. The negative-energy label is converted to the physical positron label by charge conjugation. The field expansion can be written as \cite{aleksandrov2024kinetic}
\stepcounter{equation}
\begin{widetext}
\begin{equation}\label{A2}
\begin{aligned}
\Psi(\mathbf x,t)=\sum_s\int\frac{d^3q}{(2\pi)^3}e^{i\mathbf q\cdot\mathbf x}
\bigl[&a_{\mathbf q,s}(t)u_s(\mathbf p(t))+b^\dagger_{-\mathbf q,s}(t)v_s(\mathbf p(t))\bigr].
\end{aligned}\tag{A2}
\end{equation}
\end{widetext}
The operator time dependence compensates the basis time dependence. An exact solution originating in the initial negative-energy state with label $s$ can equivalently be expanded as
\stepcounter{equation}
\begin{equation}\label{A3}
\begin{aligned}
\chi^-_{\mathbf q,s}(t)=\sum_r\bigl[
&u_r(\mathbf p(t))A^{(1)}_{rs}(\mathbf q,t)+v_r(\mathbf p(t))A^{(2)}_{rs}(\mathbf q,t)\bigr].
\end{aligned}\tag{A3}
\end{equation}
The solution originating from the initial positive-energy subspace can be written as
\stepcounter{equation}
\begin{equation}\label{A4}
\begin{aligned}
\chi^+_{\mathbf q,s}(t)=\sum_r\bigl[
&u_r(\mathbf p(t))A^{(3)}_{rs}(\mathbf q,t)+v_r(\mathbf p(t))A^{(4)}_{rs}(\mathbf q,t)\bigr].
\end{aligned}\tag{A4}
\end{equation}
At the initial time, $A^{(1)}=A^{(4)}=0$ and $A^{(2)}=A^{(3)}=I_2$. At the final time, the off-diagonal blocks encode transitions between the negative- and positive-energy subspaces. The instantaneous annihilation operator has the spin-matrix Bogoliubov form
\stepcounter{equation}
\begin{equation}\label{A5}
\begin{aligned}
a_{\mathbf q}(t)=\alpha_{\mathbf q}(t)a_{\mathbf q}^{\mathrm{in}}
+\beta_{\mathbf q}(t)b_{-\mathbf q}^{\mathrm{in}\dagger}.
\end{aligned}\tag{A5}
\end{equation}
Here $\alpha_{\mathbf q}$ and $\beta_{\mathbf q}$ are $2\times2$ matrices in spin space. The electron density matrix from vacuum initial data is $\rho_e=\beta\beta^\dagger$, while $\alpha\alpha^\dagger+\beta\beta^\dagger=I_2$. These matrix relations preserve the fermionic anticommutation relations and bound the occupation eigenvalues.

Substitution of Eq.~(\ref{A3}) into the Dirac equation and projection onto the instantaneous positive- and negative-energy subspaces gives
\stepcounter{equation}
\begin{equation}\label{A6}
\begin{aligned}
i\dot A^{(1)}+iM^{uu}A^{(1)}+iM^{uv}A^{(2)}&=\Omega A^{(1)},\\
i\dot A^{(2)}+iM^{vu}A^{(1)}+iM^{vv}A^{(2)}&=-\Omega A^{(2)}.
\end{aligned}\tag{A6}
\end{equation}
The connection matrices are overlaps of time derivatives of the instantaneous spinors. For example,
\stepcounter{equation}
\begin{equation}\label{A7}
\begin{aligned}
M^{uu}_{rs}=u_r^\dagger\dot u_s
=e\mathbf E\cdot\bigl[u_r^\dagger\nabla_{\mathbf p}u_s\bigr].
\end{aligned}\tag{A7}
\end{equation}
The second equality follows from $\dot{\mathbf p}=e\mathbf E$, not from an equality between momentum and electric field. The matrices $M^{uu}$ and $M^{vv}$ describe basis rotation within an energy subspace; $M^{uv}$ and $M^{vu}$ mix the two subspaces. For the explicitly stated spinors, these connections are~\cite{aleksandrov2024kinetic,amat2025spin}
\stepcounter{equation}
\begin{equation}\label{A8}
\begin{aligned}
M^{uu}=M^{vv}&=i\boldsymbol\mu_1\cdot\boldsymbol\sigma,\\
M^{uv}=-M^{vu}&=\boldsymbol\mu_2\cdot\boldsymbol\sigma,
\end{aligned}\tag{A8}
\end{equation}
\stepcounter{equation}
\begin{equation}\label{A9}
\begin{aligned}
\boldsymbol\mu_1=\frac{e(\mathbf p\times\mathbf E)}{2\Omega(\Omega+m)},
\end{aligned}\tag{A9}
\end{equation}
\stepcounter{equation}
\begin{equation}\label{A10}
\begin{aligned}
\boldsymbol\mu_2=
\frac{e\bigl[(\mathbf p\cdot\mathbf E)\mathbf p-
\Omega(\Omega+m)\mathbf E\bigr]}{2\Omega^2(\Omega+m)}.
\end{aligned}\tag{A10}
\end{equation}

The vector $\boldsymbol\mu_1$ generates an $SU(2)$ rotation of the instantaneous spin basis, and $\boldsymbol\mu_2$ couples the positive- and negative-energy sectors~\cite{aleksandrov2024kinetic}. Both have dimensions of inverse time. A rotating field can change both their magnitudes and directions; dropping $\boldsymbol\mu_1$ is an additional approximation and is not part of the QKE system below.

A $2\times2$ Hermitian matrix can be decomposed into a scalar and a Pauli vector. Define the electron occupation matrix by $\rho_e=F I_2-\mathbf F\cdot\boldsymbol\sigma$, so $\mathrm{tr}\rho_e=2F$. Let $\mathbf U$ and $\mathbf V$ be the real coherence vectors in the convention associated with these spinors and with the DHW transformation below. For charge-symmetric vacuum initial data, the remaining scalar coherence and charge-difference components vanish in a homogeneous electric background~\cite{aleksandrov2024kinetic}. The resulting ten real QKE variables obey~\cite{amat2025spin}
\stepcounter{equation}
\begin{equation}\label{A11}
\dot F=-2\boldsymbol\mu_2\cdot\mathbf U,\tag{A11}
\end{equation}
\stepcounter{equation}
\begin{equation}\label{A12}
\dot{\mathbf F}=2\boldsymbol\mu_1\times\mathbf F
-2\boldsymbol\mu_2\times\mathbf V,\tag{A12}
\end{equation}
\stepcounter{equation}
\begin{equation}\label{A13}
\dot{\mathbf U}=2\boldsymbol\mu_1\times\mathbf U
+\boldsymbol\mu_2(2F-1)+2\Omega\mathbf V,\tag{A13}
\end{equation}
\stepcounter{equation}
\begin{equation}\label{A14}
\dot{\mathbf V}=2\boldsymbol\mu_1\times\mathbf V
-2\boldsymbol\mu_2\times\mathbf F-2\Omega\mathbf U.\tag{A14}
\end{equation}
The scalar $F$ and the three vectors depend on $\mathbf q$ and $t$, with vacuum initial conditions
\stepcounter{equation}
\begin{equation}\label{A15}
\begin{aligned}
F(\mathbf q,t_{\mathrm{in}})&=0,\\
\mathbf F(\mathbf q,t_{\mathrm{in}})
=\mathbf U(\mathbf q,t_{\mathrm{in}})
=\mathbf V(\mathbf q,t_{\mathrm{in}})&=\mathbf0.
\end{aligned}\tag{A15}
\end{equation}

Eq.~\eqref{A11} gives the occupation source through $\boldsymbol\mu_2\cdot\mathbf U$. In Eq.~\eqref{A12}, $\boldsymbol\mu_1$ rotates the spin-density vector while $\boldsymbol\mu_2$ couples it to $\mathbf V$. Eqs.~\eqref{A13} and \eqref{A14} evolve the two coherence quadratures, with the energy gap $2\Omega$. The factor $2F-1$ encodes fermionic occupation feedback. Pauli constraints apply to the eigenvalues of the full density matrix, not only its scalar part.

The system possesses the exact invariant
\stepcounter{equation}
\begin{equation}\label{A16}
	\left( \frac{1}{4} \right) (1 - 2F)^2 + \left| \vec{F} \right|^2 + \left| \vec{U} \right|^2 + \left| \vec{V} \right|^2 = \frac{1}{4}. \tag{A16}
\end{equation}
For Eqs.~\eqref{A11}--\eqref{A14}, Eq.~\eqref{A16} is conserved and implies $0\leq F\leq1$. The stronger physical occupation bounds are $0\leq F\pm|\mathbf F|\leq1$, as follows from the fermionic density matrix. Monitoring the invariant is a useful integration check~\cite{aleksandrov2024kinetic}, but an invariant alone does not prove convergence of a momentum spectrum or agreement of spin projectors.

In the asymptotic basis specified above, the electron occupations for the rest-frame spin labels $s=\pm1$ are
\stepcounter{equation}
\begin{equation}\label{A17}
f_s^{(e^-)}(\mathbf q)=F(\mathbf q,t_{\mathrm{out}})
-sF_z(\mathbf q,t_{\mathrm{out}}).\tag{A17}
\end{equation}

With charge-conjugate physical positron labels and opposite canonical momentum, the positron occupations are
\stepcounter{equation}

\begin{equation}\label{A18}
f_s^{(e^+)}(\mathbf q)=F(-\mathbf q,t_{\mathrm{out}})
+sF_z(-\mathbf q,t_{\mathrm{out}}).\tag{A18}
\end{equation}

The two electron channels sum to $2F$. In the main text, $f_{\mathrm{DHW}}$ is also spin summed and satisfies $f_{\mathrm{up}}+f_{\mathrm{down}}=f_{\mathrm{DHW}}$. Matching the density matrices therefore gives
\stepcounter{equation}

\begin{equation}\label{A19}
f_{\mathrm{DHW}}(\mathbf q,t)=2F(\mathbf q,t).\tag{A19}
\end{equation}

The vector $\mathbf F$ changes under spin-basis rotations, while $\mathrm{tr}\rho_e$ is basis invariant. Eq.~\eqref{A17} gives the specified boosted rest-frame spin projection, and an arbitrary direction is obtained as $F\mp\mathbf n\cdot\mathbf F$. A helicity measurement uses the direction of the final kinetic momentum; for electrons and $|\mathbf p|>0$, the left- and right-helicity distributions are
\stepcounter{equation}

\begin{equation}\label{A20}
\begin{aligned}
f_{L/R}^{(e^-)}(\mathbf q,t)=F(\mathbf q,t)
\pm\frac{\mathbf p(\mathbf q,t)\cdot\mathbf F(\mathbf q,t)}
{|\mathbf p(\mathbf q,t)|}.
\end{aligned}\tag{A20}
\end{equation}

The upper sign in Eq.~\eqref{A20} is left helicity and the lower sign is right helicity. Helicity is undefined at zero momentum. The $z$-axis spin projection coincides with helicity only when the momentum lies on that axis, with the appropriate orientation. For massive particles, helicity should also be distinguished from chirality.

For a field $\mathbf E(t)=E(t)\mathbf n_0$ with fixed direction $\mathbf n_0$, the vacuum evolution has $\mathbf F=\mathbf0$ and can be represented by three real functions in an adapted basis~\cite{hebenstreit2010relating,aleksandrov2024kinetic}. This is not a statement that the scalar occupation has zero derivative. The scalar coherence variables $\mathcal U$ and $\mathcal V$ below are the adapted quantum-Vlasov variables; for nonzero transverse momentum they are not simply fixed-direction components of the general vectors $\mathbf U$ and $\mathbf V$. The reduced equations are
\stepcounter{equation}

\begin{equation}\label{A21}
\begin{aligned}
\dot F&=-2\mu_2\mathcal U,\\
\dot{\mathcal U}&=\mu_2(2F-1)+2\Omega\mathcal V,\\
\dot{\mathcal V}&=-2\Omega\mathcal U,
\end{aligned}\tag{A21}
\end{equation}
where
\stepcounter{equation}

\begin{equation}\label{A22}
\begin{aligned}
\mu_2=\frac{eE(t)\varepsilon_\perp}{2\Omega^2(\mathbf q,t)},
\quad\varepsilon_\perp=\sqrt{m^2+q_\perp^2}.
\end{aligned}\tag{A22}
\end{equation}

Eq.~\eqref{A21} is the standard fixed-direction quantum-Vlasov system~\cite{kluger1992fermion,hebenstreit2010relating}. Here $q_\perp$ is transverse to $\mathbf n_0$, in a gauge with no transverse constant potential, and the scalar $\mu_2$ is defined by Eq.~\eqref{A22} in the adapted coherence convention. Setting $C=\mathcal U+i\mathcal V$ and integrating $\dot C=\mu_2(2F-1)-2i\Omega C$ with vacuum initial data yields the non-Markovian equation for the occupation rate~\cite{hebenstreit2010relating,kluger1998quantum}:
\stepcounter{equation}

\begin{widetext}
\begin{equation}\label{A23}
\begin{aligned}
\dot F(\mathbf q,t)={}&2\mu_2(t)\int_{t_{\mathrm{in}}}^{t}dt'\,
\mu_2(t')\bigl[1-2F(\mathbf q,t')\bigr]\cos\!\left[2\int_{t'}^{t}d\tau\,\Omega(\mathbf q,\tau)\right].
\end{aligned}\tag{A23}
\end{equation}
\end{widetext}

The memory integral makes the production rate depend on the earlier field history. Its cosine contains the gap phase accumulated between two times, and $1-2F$ is the fermionic inversion factor~\cite{dumlu2010stokes,hebenstreit2010relating}. The three-component reduction applies to the fixed-direction electric-field problem with the stated initial data. A generic rotating field requires retaining the spin-density and coherence vectors in Eqs.~\eqref{A11}--\eqref{A14}; special parameter or momentum symmetries may simplify particular modes, but are not assumed here.
\begin{figure*}
	\centering
	\includegraphics[width=1\linewidth]{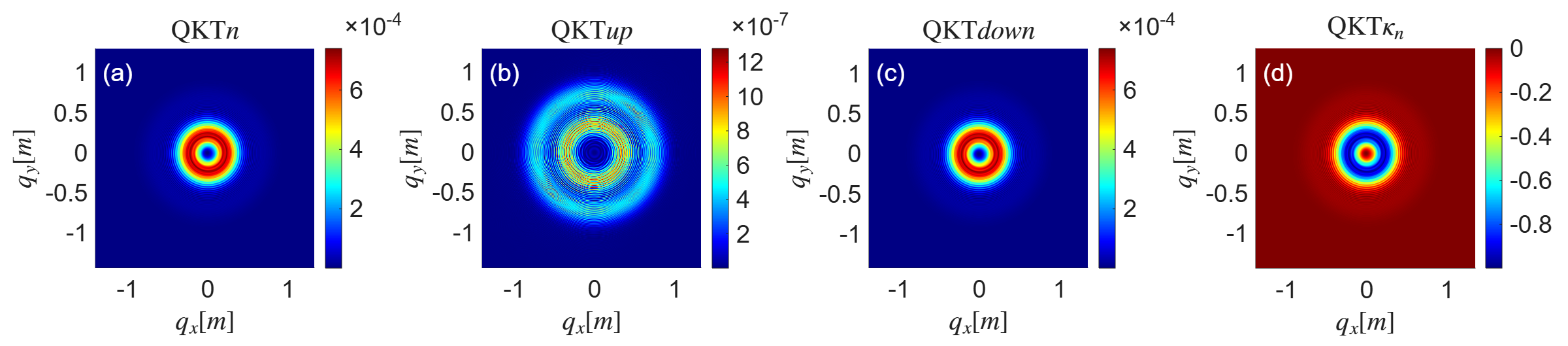}
	\caption{QKT electron spectra and normalized spin difference for high-frequency circular polarization ($\delta=1$) on $q_z=0$. Columns show $f=2F$ (labeled QKT$n$), $f_{\mathrm{up}}$, $f_{\mathrm{down}}$, and $\kappa_n$. Field strength, carrier frequency, and duration are as in Fig.~\ref{fig3}. A quantitative DHW comparison requires the same chirp, momentum convention, and spin projector; the channel color bars have different scales.}
	\label{fig11}
\end{figure*}

\begin{figure*}
	\centering
	\includegraphics[width=1\linewidth]{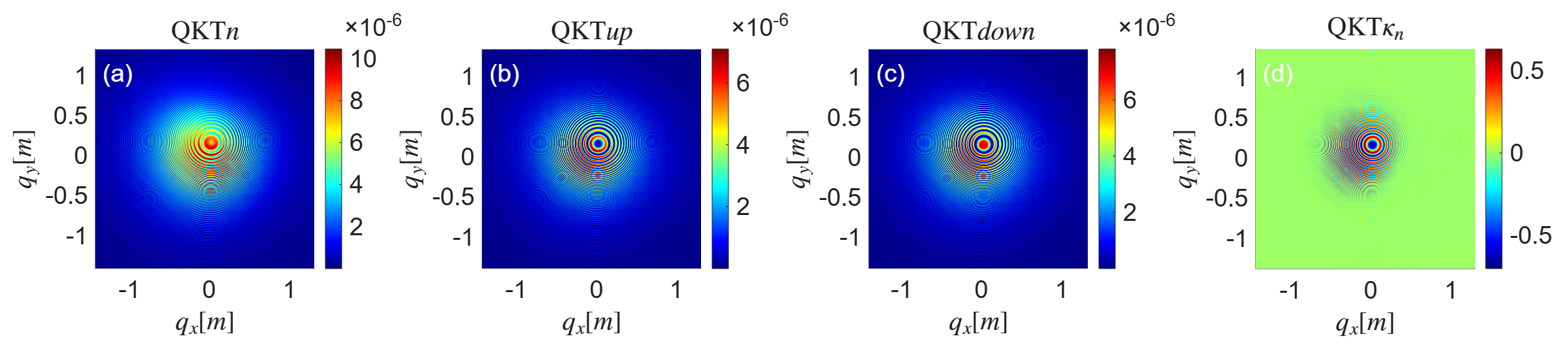}
	\caption{QKT electron spectra and normalized spin difference for low-frequency circular polarization ($\delta=1$) on $q_z=0$. Columns show $f=2F$ (labeled QKT$n$), $f_{\mathrm{up}}$, $f_{\mathrm{down}}$, and $\kappa_n$. Field strength, carrier frequency, and duration are as in Fig.~\ref{fig8}. A quantitative DHW comparison requires the same chirp, momentum convention, and spin projector; the channel color bars have different scales.}
	\label{fig12}
\end{figure*}
Let $s_D$, $\mathbf v_D$, $\mathbf a_D$, and $\mathbf t_D$ denote the raw equal-time DHW components, rather than the auxiliary vector $\mathbf v$ in Eq.~\eqref{20}. Their instantaneous vacuum values are $s_{\mathrm{vac}}=-2m/\Omega$ and $\mathbf v_{\mathrm{vac}}=-2\mathbf p/\Omega$~\cite{bialynicki1991phasespace,vasak1987quantum,hebenstreit2010relating}. With the spinor and tensor conventions stated above, the forward transformation is

\stepcounter{equation}
\begin{equation}\label{A24}
s_D-s_{\mathrm{vac}}=\frac4\Omega
\bigl[mF-\mathbf p\cdot\mathbf U\bigr].\tag{A24}
\end{equation}

\stepcounter{equation}
\begin{equation}\label{A25}
\begin{aligned}
\mathbf v_D-\mathbf v_{\mathrm{vac}}=\frac4\Omega\biggl[
&\Omega\mathbf U+\mathbf pF-\frac{\mathbf p(\mathbf p\cdot\mathbf U)}{\Omega+m}\biggr].
\end{aligned}\tag{A25}
\end{equation}

\stepcounter{equation}

\begin{equation}\label{A26}
\begin{aligned}
\mathbf a_D=-\frac4\Omega\biggl[m\mathbf F-\mathbf p\times\mathbf V
+\frac{\mathbf p(\mathbf p\cdot\mathbf F)}{\Omega+m}\biggr].
\end{aligned}\tag{A26}
\end{equation}

\stepcounter{equation}

\begin{equation}\label{A27}
\begin{aligned}
\mathbf t_D=-\frac4\Omega\biggl[m\mathbf V-\mathbf p\times\mathbf F
+\frac{\mathbf p(\mathbf p\cdot\mathbf V)}{\Omega+m}\biggr].
\end{aligned}\tag{A27}
\end{equation}

The inverse relations are
\stepcounter{equation}

\begin{equation}\label{A28}
\begin{aligned}
\mathbf U=-\frac1{4\Omega}\biggl[\mathbf p s_D-\Omega\mathbf v_D
+\frac{\mathbf p(\mathbf p\cdot\mathbf v_D)}{\Omega+m}\biggr].
\end{aligned}\tag{A28}
\end{equation}

\stepcounter{equation}

\begin{equation}\label{A29}
\begin{aligned}
\mathbf F=-\frac1{4\Omega}\biggl[m\mathbf a_D+\mathbf p\times\mathbf t_D
+\frac{\mathbf p(\mathbf p\cdot\mathbf a_D)}{\Omega+m}\biggr].
\end{aligned}\tag{A29}
\end{equation}

\stepcounter{equation}

\begin{equation}\label{A30}
\begin{aligned}
\mathbf V=-\frac1{4\Omega}\biggl[m\mathbf t_D+\mathbf p\times\mathbf a_D
+\frac{\mathbf p(\mathbf p\cdot\mathbf t_D)}{\Omega+m}\biggr].
\end{aligned}\tag{A30}
\end{equation}

Eqs.~\eqref{A24}--\eqref{A30}, together with $F=[m(s_D-s_{\mathrm{vac}})+\mathbf p\cdot(\mathbf v_D-\mathbf v_{\mathrm{vac}})]/(4\Omega)$, form an algebraically invertible transformation. In particular, the minus sign of the longitudinal coherence term in Eq.~\eqref{A25} is required to recover $f_{\mathrm{DHW}}=2F$. With identical external field, initial state, momentum convention, and tensor normalization, the two systems describe the same homogeneous-field quantum evolution~\cite{hebenstreit2010relating,aleksandrov2024kinetic}. Comparing spin-resolved spectra additionally requires identical final-state projectors. Thus agreement between implementations is a useful numerical check, rather than evidence from independent physical approximations.

We did this work using the QKT method, and the results are consistent with the DHW method.Below are the momentum spectrum and the asymmetrical distribution for circular polarization. Figs.~\ref{fig11} and ~\ref{fig12} show the cases of circular polarization at high and low frequencies, respectively.

\FloatBarrier           
\bibliography{refs}
\end{document}